\documentclass[pre,twocolumn,aps,amssymb,footinbib,showpacs]{revtex4}
\usepackage{amssymb}
\usepackage{amsmath}
\usepackage{times}
\usepackage{latexsym}
\usepackage{graphicx}
\usepackage{color}
\usepackage{bm}
\usepackage{subfigure}

\begin{document}
\title{ Exploring complex phenomena using ultracold atoms in bichromatic lattices }
\author{ Shuming Li$^{1}$,  Indubala I Satija$^{2,3}$,  Charles W. Clark$^{3}$ and  Ana Maria Rey$^{1}$}
\affiliation{$^{1}$  JILA, NIST, Department of Physics, University of Colorado, 440, UCB, Boulder, CO, 80309}
 \affiliation{$^{2}$ Department  of Physics, George Mason University, Fairfax, VA, 22030,USA}
\affiliation{ $^{3}$ Joint Quantum Institute, National Institute of Standards and
Technology and University of Maryland, Gaithersburg MD, 20899, USA}

\date{\today}
\begin{abstract}

With an underlying common theme of competing length scales,
we study the many-body Schr\"{o}dinger equation in a quasiperiodic potential
and discuss its connection with the Kolmogorov-Arnold-Moser (KAM) problem of classical mechanics.
We propose a possible visualization of such connection in experimentally accessible many-body observables.
Those observables are useful probes for the three characteristic phases of the problem:
 the metallic, Anderson and band insulator phases. In addition, they exhibit fingerprints of non-linear phenomena such as
Arnold tongues, bifurcations and devil's staircases.
Our numerical treatment is complemented with a perturbative analysis which provides insight on the underlying physics.
The perturbation theory approach is particularly useful in illuminating the distinction between the Anderson insulator and the band insulator phases in terms of paired sets of dimerized states.
\end{abstract}
\maketitle

\section{Introduction}

Ultracold atoms are emerging as a versatile arena for the study of a variety of problems in physics.
These clean and highly controllable systems can be viewed as  simulators of complex quantum phenomena
with applications in condensed matter, quantum optics, atomic and molecular physics and nonlinear dynamics, as well as in particle physics and cosmology \cite{Bloch2008r, Lewenstein2007}.
Some examples of this trend are the laboratory realization of the superfluid to Mott insulator
 transition with bosons \cite{Greiner2002a}, observation of the corresponding metal to Mott insulator transition with fermions \cite{Jordens2008, Schneider2008} and the creation of the Tonks-Girardeau gas in one dimensional systems \cite{Kinoshita2004,Paredes2004}.
Ultracold atoms have also provided a laboratory realization of Wannier-Stark ladders \cite{Wilkinson1996} and the
kicked-rotor model \cite{Moore1995}, a paradigm in the study of classical and quantum chaos
\cite{Moore1994}. Recent successful loading of atoms
in a bichromatic optical lattice geometries \cite{Fallani2007}, has opened new avenues
for the study of phenomena where the competition between various length scales is at the heart of the problem.

Systems with competing lengths have fascinated physicists as well as mathematicians
in view of their exotic fractal characteristics \cite{Hofstadter1976,Azbel1983}. Such systems with two competing periodicities,
commonly known as almost-periodic or quasiperiodic, occur very commonly in nature.
The most commonly studied example in quantum physics is  the single particle Schr\"{o}dinger equation in the presence of a quasiperiodic (QP) potential.

Motivated by experimental realization of two-color lattices, we revisit the problem of the Schr\"{o}dinger equation in a quasiperiodic potential
and its relationship with the Kolmogorov-Arnold-Moser (KAM) \cite{Kolmogorov1953} problem of classical mechanics.  Our particular focus is the treatment of many-body effects that are present in QP systems that have been extensively studied at the single-particle level. By focusing on experimentally accessible observables such as
the momentum distribution and density-density correlations, we demonstrate the possibility of experimental
visualization of the relationship between the metal-insulator transition and the KAM-Cantori transition \cite{Percival1979} in many body systems.
These observables are found to exhibit fingerprints of various paradigms of
nonlinear systems such as Arnold tongues, bifurcations and devil's staircases. One of the aims of this paper is to
communicate the excitement of ultracold atomic physics to the nonlinear dynamics community. We also hope that cold atom community will also
benefit from our discussion of the relationship between problems of  condensed matter theory and nonlinear dynamics.

The many-body systems that we treat here are ensembles of ultracold spin-polarized fermionic atoms confined in one-dimensional bichromatic optical lattices. We treat cases in which the two lattices have incommensurate periodicities, and thus constitute a quasi-periodic potential for atomic motion \cite{Drese1997}.
Since the metal-insulator transition in single-particle QP systems is associated with a localization transition of extended single-particle states,
a natural probe of such transitions in many-particle systems are the density distribution and density-density correlations.
At present, the density distribution of atoms confined in a lattice has not been easily accessible to experimental measurement, instead
the quasi-momentum distribution and its corresponding correlation functions, have been measured after the lattice potential has been suddenly removed.

Using  the momentum-position duality of our basic model \cite{Aubry1979,Sokoloff1985},
we show that time of flight images  encode local density information and can be used to identify the possible phases of ultracold atomic gases, e.g., metallic phases and various types of insulators. 
Many of our results are explained analytically using perturbation theory, although the complete picture is  obtained by exact numerical calculations, particularly near the metal-Anderson insulator transition.

The paper is organized as follows.
Section II contains an overview of experiments on ultracold atoms in optical lattices. Section III describes the basic Hamiltonian, and the experimental
observables.  In section IV, we describe the relationship between
the metal-Anderson insulator transition  and KAM-Cantori transition. In Section V we discuss possible experimental manifestations of effectively nonlinear behavior in these many-body systems.  Sec. VI provides a summary of our results and states our conclusions.
\section{ Ultracold Atoms in Optical Lattices}

An optical lattice is created by the interference of counter-propagating laser beams that give rise to a spatially periodic intensity pattern.
The intensity pattern corresponds directly to a potential for neutral atoms via the a.c. Stark shift of the atomic energy levels.
The two important parameters of an optical lattice are the depth of the lattice potential wells and the lattice constant, $a$. The well depth of the lattice can be tuned
by changing the intensity of the laser, while
$a$ can be tuned by changing the wavelength of the laser or by changing the relative angle between the two laser beams \cite{Bloch2008r}.

The systems of interest here are gases of ultra-cold spin-polarized  fermionic atoms trapped in the lowest band of a transversal 2D optical lattice. For most such systems of current experimental interest, where atomic interactions are short-range, spin-polarized fermions are effectively noninteracting due to Pauli blocking. The 2D lattice depth is made strong enough to freeze the motion of the atoms transversally, creating an array of independent 1D
tubes  \cite{Kinoshita2004,Tolra2004a}. Along the axis of the 1D tubes, additional optical lattices can be imposed, as has been done in Refs. \cite{Fertig2005,Paredes2004}.  We discuss cases in which two such lattices are imposed, with incommensurate periods \cite{Lewenstein2007} (when one of these lattices is much stronger than the other, we refer to it as the primary lattice). 
The combined lattices therefore generate an effective quasi-periodic (QP) potential along the axial direction. Such an experiment has been recently implemented for bosonic atoms in Ref. \cite{Fallani2007}.
Our treatment
considers cases where the ratio of the two lattice constants is equal to the ``golden mean'', $\sigma =(\sqrt{5}-1)/2$, which is one of the best-studied examples in single-particle physics \cite{Sokoloff1985}.

In most experiments atoms are first trapped and cooled to quantum degeneracy.  They are subsequently loaded into the lattice
by adiabatically turning on the lattice laser beams. At the end of each experimental sequence atoms are probed by using time of flight images. These are obtained after releasing the atoms by turning off all the confinement potentials.   The atomic cloud expands and then photographed after it enters the ballistic regime.
Assuming that the atoms are
noninteracting from the time of release, properties of the initial
state can be inferred from the spatial images
\cite{Altamn2004}: the column density distribution image
reflects the initial  quasi-momentum distribution,  and the
density fluctuations, namely the noise correlations, reflect
the quasi-momentum fluctuations. These quantities, which  will be defined below --see Eqs. (\ref{nnoise1},\ref{nnoise})--,
have been shown to be successful diagnostic tools for characterizing quantum phases and have been recently measured in  bosonic  quasi-periodic systems \cite{Guarrera2008}.

\section{ Our Model System: The Harper Equation, Many-body  Observables and self-Duality }

\subsection{The Harper Equation }
If the intensity of the secondary lattice is much weaker than
that of the primary lattice,  the low energy physics of the fermionic system  can be well described
 by the tight-binding Hamiltonian \cite{Lewenstein2007}:
\begin{equation}
H=-J\sum_j(\hat{c}_j^{\dagger}\hat{c}_{j+1}
+\hat{c}_{j+1}^{\dagger}\hat{c}_{j})+\sum_{j} V_j
\hat{n}_j,\label{fermion}
\end{equation}
where $\hat{c}_j$ is the fermionic annihilation operator at the lattice  site
$j$, and $J$ is the  hopping energy
between adjacent sites.  The main effect of the QP potential
 is to modulate the on site potential.  It is accounted for by the term  $V_j=2 V_0 \cos( 2\pi \sigma j+\phi )$.
The parameter $V_0$ is proportional to the intensity of the lasers
used to create the secondary lattice \cite{Drese1997}, $\sigma$ is the ratio between the wave vectors of the two lattices which  we choose to take value $\sigma=(\sqrt{5}-1)/2$, and $\phi$ is a phase factor that is determined by the absolute registration of the two lattices.

 To model real experimental conditions, averaging over $\phi$ is required. This averaging takes into account,
on one hand, the  phase fluctuations  from one preparation to
another. Those arise due to the difficulty
to lock the position of the cloud over several
shots. On the other, the fact that typical experimental set-ups
generally consist of an assembly of one-dimensional  tubes  with different lengths and  phases among them.

For a single atom, the eigenfunctions  $\psi_j^{(m)}$  and
eigenenergies $E^{(m)}$  of  the Hamiltonian in  Eq.(\ref{fermion})
satisfy:
\begin{equation}
-(\psi_{j+1}^{(m)}+\psi_{j-1}^{(m)})+  2 \lambda \cos(2\pi\sigma j+\phi )
\psi_{j}^{(m)}=\epsilon^{(m)} \psi_{j}^{(m)}.\label{harpereqn}
\end{equation}
Where $\lambda=V_0/J$, $\epsilon^{(m)}=E^{(m)}/J$, and $\epsilon^{(m)}\leq\epsilon^{(m+1)}$.  Eq. (\ref{harpereqn}) is known as the Harper equation, a paradigm in the study of 1D
quasiperiodic (QP) systems \cite{Sokoloff1985}. For irrational $\sigma$, the model
exhibits a transition from extended to localized states at
$\lambda_c=1$. Below criticality,
all the states are extended Bloch-like states characteristic of a periodic potential. Above criticality the Harper model
becomes equivalent to a corresponding Anderson model, the spectrum
is a pure point spectrum and all  states are exponentially
localized. At criticality the spectrum is a Cantor set and the gaps
form a devil's staircase  of measure unity \cite{Hofstadter1976}.

In our numerical studies, $\sigma$ is approximated by the ratio of two Fibonacci
numbers $F_{M-1}/F_{M}$, ($F_1=F_0=1, F_{i+1}=F_{i}+F_{i-1}$), which
describe the best rational approximant by continued fraction
expansion of $\sigma$. For this rational approximation the unit cell
has length $F_M$ and the single-particle spectrum consists of $F_M$
bands and $F_{M}-1$ gaps. The gaps occur at  $Q_{n}/2$,  $\pm(\pi-|Q_{n}|/2)$ with $Q_{n}=\pm(2\pi/a)\langle n\sigma\rangle$
 reciprocal
lattice vectors  constrained in the interval  $a Q_{n} \in(-\pi,\pi]$. Here  $\langle n \sigma\rangle = n \sigma$ (mod 1), $n$ an integer .
We denote by  $N_p$ the number of atoms, $N_l=F_M$ is the number of lattice sites, and the filling factor $\nu=N_p/N_l$ ranges from 0 to 1.

\subsection{Many-body  Observables }
An ensemble of spin-polarized fermions at zero temperature are ``stacked up" into the single-particle eigenstates of increasing energy,
with one particle per quantum state. The energy of the highest occupied level,  which depends on  the filling factor $\nu$, is the Fermi energy, $E_F$. Since at the critical point  all the single-particle wave functions become localized, at the many-body level polarized fermions
also exhibit a transition from metal to
insulator at $\lambda_c$. However, in addition to these two phases the fragmentation of the single-particle spectrum in a series of bands and gaps  introduces additional band insulator phases when the Fermi energy lies within a gap. The most relevant insulating phases occur at the irrational filling factors: $\nu=F_{M-1}/F_M$ and $\nu=F_{M-2}/F_M$ ( which respectively correspond to $\nu=\sigma,1-\sigma$), associated with the leading band gaps.
In the band insulator phases the many body system is an insulator, irrespective of the value of $\lambda$, even though extended single particle states are occupied.

Since the metal-insulator transition is clearly signaled by the onset of localization of extended single-particle states,
a natural probe of this transition is the many-body density profile, $\rho_j$,  and the density-density correlations, $\Delta(j_{1},j_{2})$, which can be written in terms of single particle wave functions as:
\begin{eqnarray}\rho_j &=& \langle \hat{n}_{j}\rangle= \sum_{m = 1}^{N_p} {\left|\psi_j^{(m)}\right|}^2\label{den}\\\Delta(j_{1},j_{2})&=&\langle \hat{n}_{j_1}\hat{n}_{j_2}\rangle-\langle \hat{n}_{j_1}\rangle \langle \hat{n}_{j_2}\rangle \notag\\ &=&\sum_{m=1}^{N_p}\left|\psi^{(m)}_{j_{1}}\right|^2\delta_{j_{1},j_{2}}-\left|\sum_{m=1}^{N_p}
{\psi_{j_{1}}^{(m)}}^{*}\psi_{j_{2}}^{(m)}\right|^2
\end{eqnarray} Here $\hat{n}_{j}=\hat{c}_{j}^{\dagger}\hat{c}_{j}$ and we have used Wick's theorem to evaluate $\Delta(j_{1},j_{2})$.
However, in general, such local observables are hard to measure experimentally due to the lack of addressability of individual lattice
sites for typical lattice spacing. Instead, time of flight images access non-local observables such as the quasi-momentum distribution,
$\hat{n}(Q)$ and
the quasi-momentum fluctuations,  $\Delta(Q,Q')$, which are given by:
\begin{eqnarray}
\hat{n}(Q)&=&\frac{1}{N_{l}}\sum_{i,j}  e^{i Q a(i-j)} \hat{c}_{i}^{\dagger}\hat{c}_{j},\label{nnoise1}
\\\Delta(Q,Q')&=&\langle\hat{n}(Q) \hat{n}(Q')\rangle-\langle\hat{n}(Q)\rangle\langle\hat{n}(Q')\rangle,
\label{nnoise}\end{eqnarray} 
where $Q$ is the quasi momentum which can be expressed in terms of the  indices $k=0,1,2,\dots$ as:
$Q(k)=\frac{2\pi}{a}\frac{k}{N_l} $. $Q(k)$ is constrained to the  interval $Qa\in(-\pi,\pi]$ .

  Introducing $\eta_k$,  the Fourier transform of $\psi_k$,
\begin{equation}
\eta_k^{(m)}=\frac{1}{\sqrt{N_l}}\sum_j e^{-i \frac{2\pi k j}{N_{l}}} \psi_j^{(m)},
\label{FT}
\end{equation} the observables   $n_{k}$ (also denoted sometimes as $n(Q(k))$) and $\Delta(Q)\equiv \Delta(Q,0)$
 can be written  as:
\begin{eqnarray}
n_{k}&=&\sum_{m=1}^{N_{p}}\left|{\eta_k^{(m)}}^{*}\eta_{k}^{(m)}\right|,\\
\Delta(Q(k))&=&\sum_{m=1}^{N_p}\left|\eta^{(m)}_0\right|^2\delta_{k,0}-\left|\sum_{m=1}^{N_p}{\eta_k^{(m)}}^{{*}}\eta_0^{(m)}\right|^2.\end{eqnarray}

\subsection{Self-duality}

The self-duality of the Harper equation corresponds to the property that single particle eigenstates $\psi_j$ and their corresponding
Fourier transformed eigenstates, $z_n\equiv\frac{1}{\sqrt{N_l}} e^{- i n \phi} \sum \psi_j e^{-i j (2\pi \sigma n +\theta)}$
satisfy the same equation with the roles of $J$ and $V_0$ interchanged \cite{Aubry1979, Sokoloff1985}:
\begin{equation}
 -(z_{n-1}+z_{n+1} )+ \frac{2}{\lambda} \cos ( 2\pi \sigma n+\theta) z_n  =- \frac{\epsilon}{\lambda} z_n.\label{harpereqnF}
\end{equation}This relation implies that the experimentally relevant variables, $\eta_k$, (see Eq. (\ref{FT}))  also satisfy
\begin{eqnarray}
 &&-(\eta_{\text{Mod}[k+F_{M-1},F_M]}e^{-
 i\phi}+\eta_{\text{Mod}[k-F_{M-1},F_M]}e^{
 i\phi} ) \nonumber \\
 &&+\frac{2}{\lambda} \cos(2\pi k/F_M) \eta_k  =- \frac{\epsilon}{\lambda} \eta_k\label{harpereqnk1}
\end{eqnarray}

As discussed below,
self-duality is a key to obtaining experimental information on local quantities from measurements. In this context,
it is important to understand the relationship between the index $j$ of $\psi_j$ that satisfies the Harper equation (\ref{harpereqn})
and its corresponding index $k$ in $\eta_k$ that satisfies the dual equation (\ref{harpereqnk1}).
An example of the mapping is provided in Appendix A.
It should be noted that Fibonacci sites in real space are mapped to Fibonacci sites in the momentum space, up to a common displacement.
This shift is dependent on the phase factor $\phi$.
We obtain this relationship numerically by diagonalizing the Harper equation for a given  $\lambda$, labeling the states in increasing order
in energy, and then finding the corresponding momentum space dual by repeating the same procedure but with $\lambda$ replaced by $1/\lambda$.

\section{Localization Transition as a KAM-Cantori transition}

The perturbative  treatment of the QP potential is fundamentally related to the treatment of a non-integrable perturbation applied to an integrable
Hamiltonian system in classical mechanics. Both cases exhibit the  small-divisors problem  related to
the presence of  high-order terms with small denominators in the perturbation
expansion.
In the classical system,   Kolmogorov, Arnold and Moser (KAM) \cite{Kolmogorov1953} solved the problem and 
demonstrated that most of the invariant tori in the phase space  are not destroyed by a sufficiently weak nonintegrable perturbation.
Outside the perturbative regime, KAM tori break, becoming an invariant cantor set, known as Cantori \cite{Percival1979}.

The study of the Harper equation is fundamentally related to the KAM type problems of classical mechanics.
Both systems share the mathematical difficulty of having  higher-order terms with small 
denominators when the quasiperiodic potential or non-integrable term  are treated perturbatively.
Under this point of view, the metallic phase in the Harper Equation, with continuous spectrum and Bloch-type wave functions has been identified as the analog of the KAM phase with invariant tori,
while the localized phase of the Harper system with point spectrum and exponentially localized states has been usually compared with the Cantori phase.
At the single-particle level this connection has been visualized by means of the so called Hull function \cite{Ostlund1983,Ostlund1983b} defined as $\psi_n= e^{i n \alpha } \chi(n \sigma)$, with $\alpha$ a real phase factor.
$\chi(n \sigma)$ is a smooth and  continuous function in the extended phase but  becomes discontinuous for $\lambda>1$. Here we propose instead to look at the return map of the local density of the atomic cloud,  $\rho_j$ vs $\rho_{j+1}$,  as a cleaner visualization of the KAM to Cantori transition
at the many-body level.

\begin{figure}[htbp]
\subfigure[]{\includegraphics[width =0.48\linewidth]{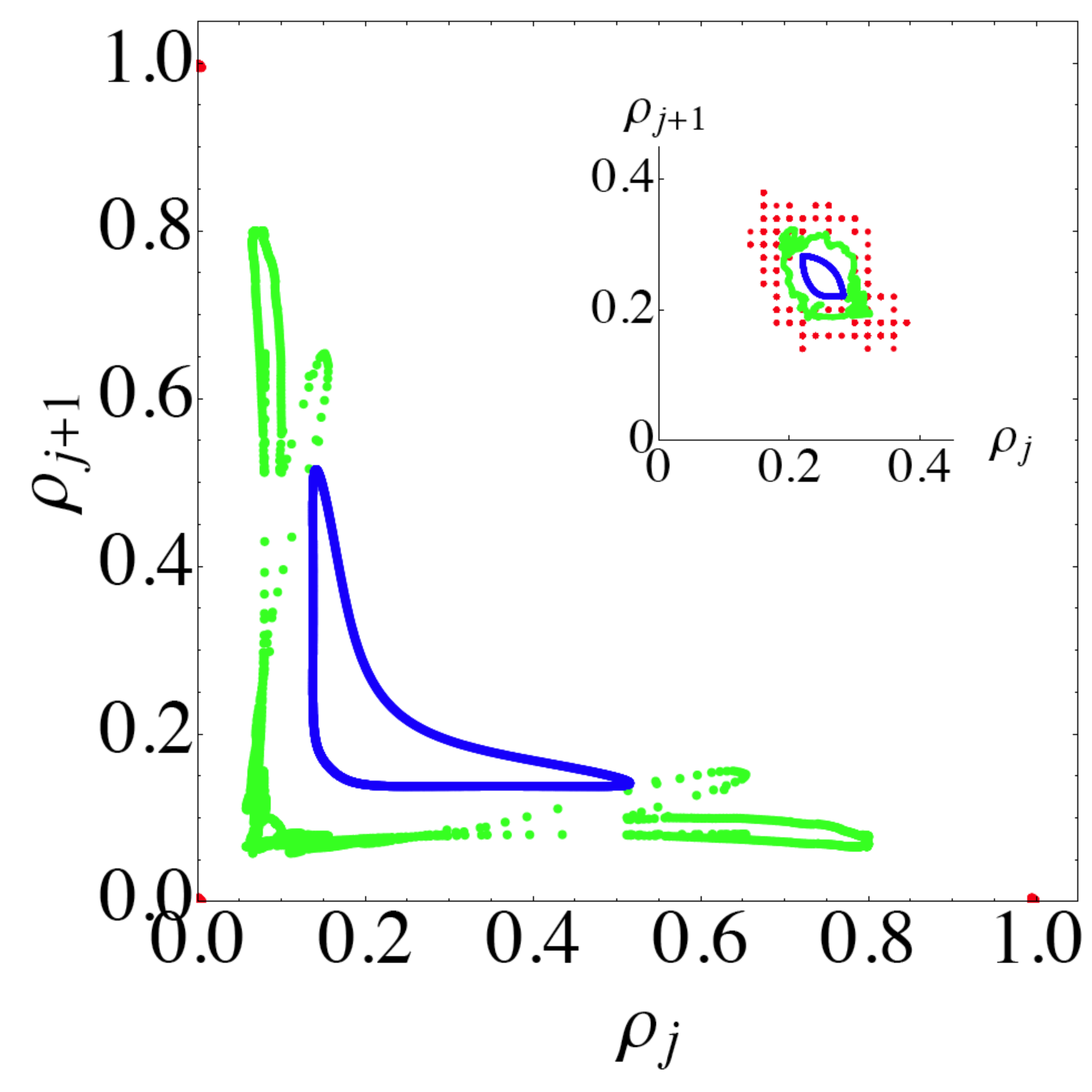}}\quad
\subfigure[]{\includegraphics[width =0.48\linewidth]{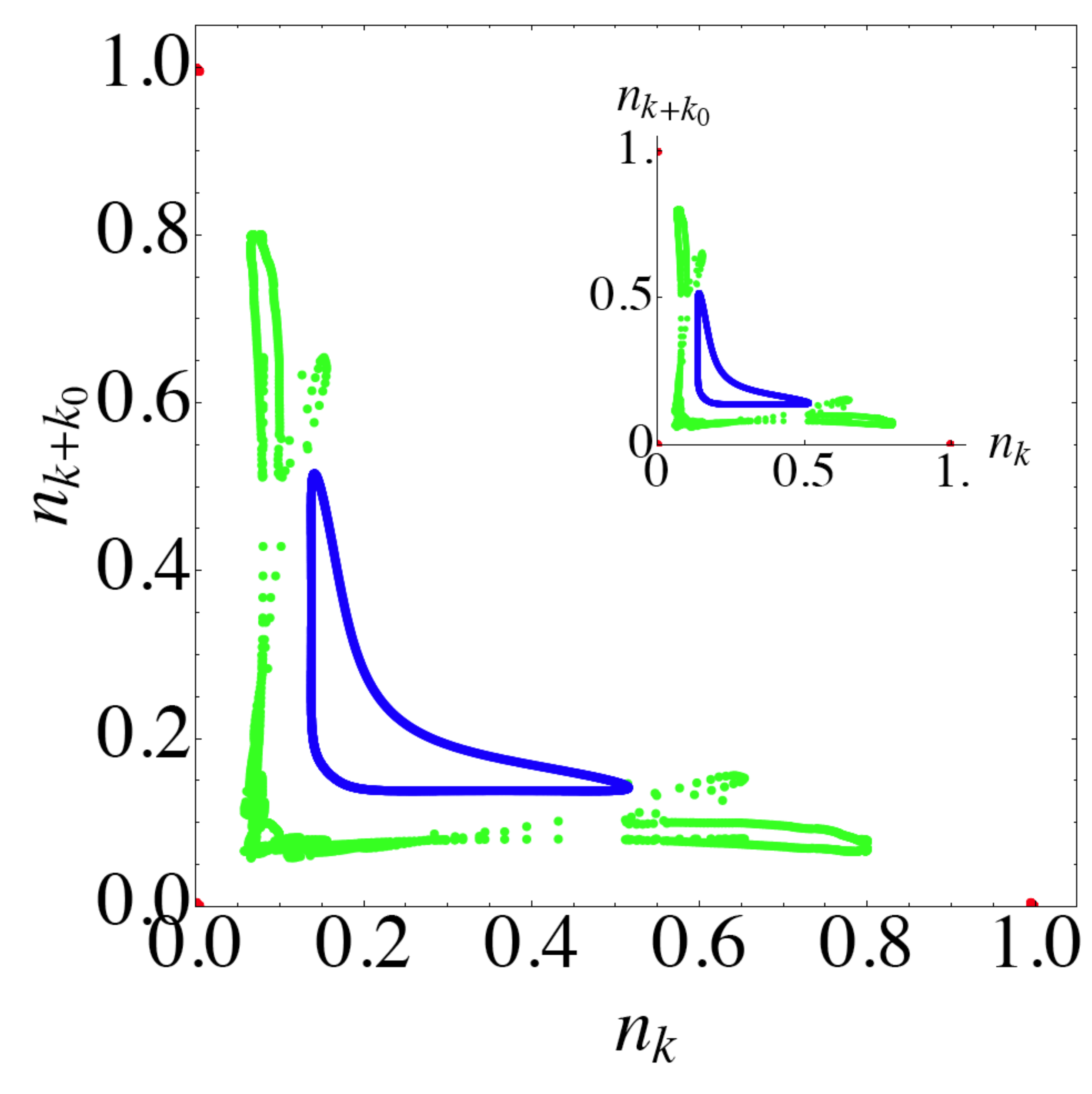}}\\
\subfigure[]{\includegraphics[width =0.48\linewidth]{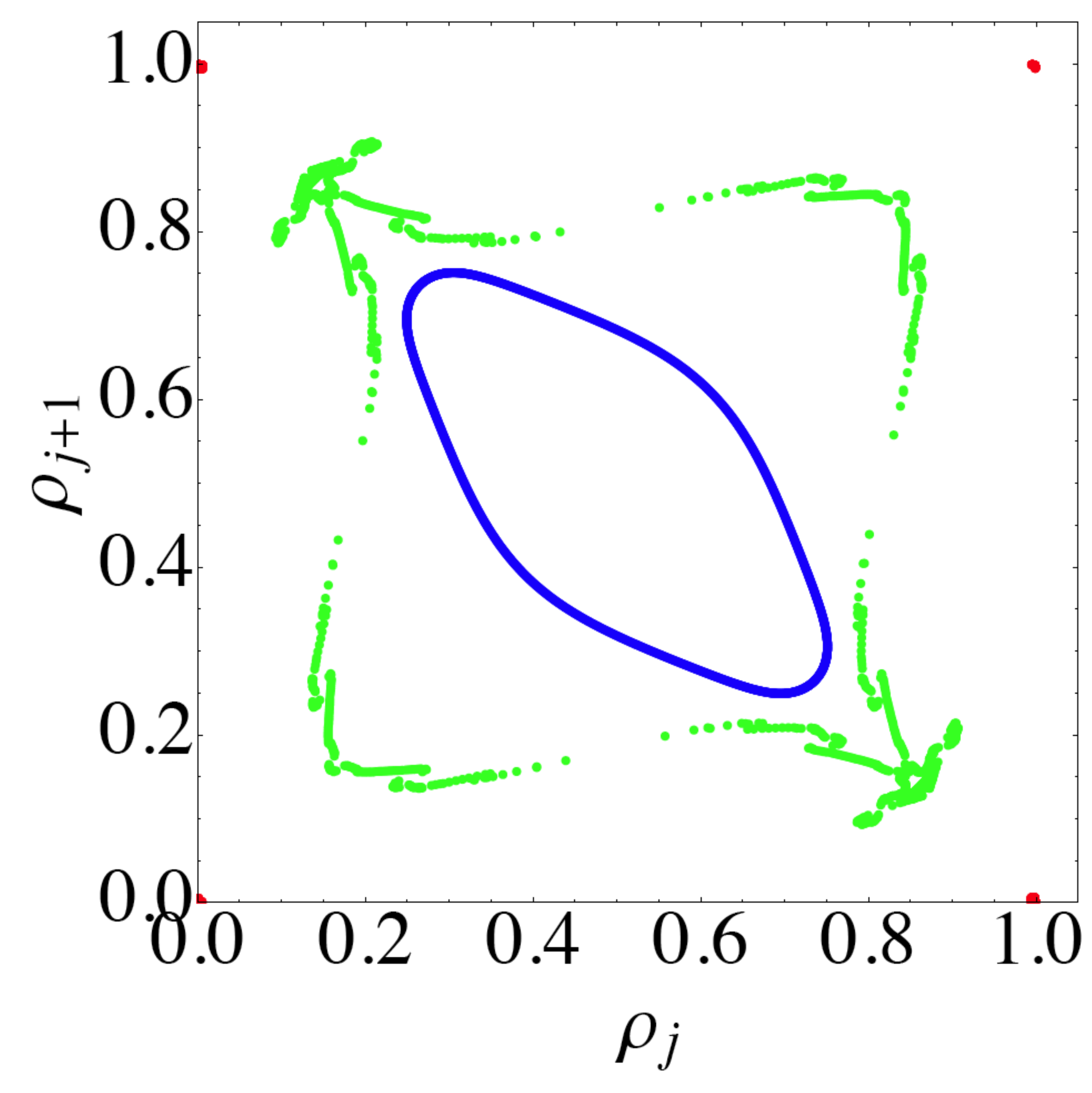}}\quad
\subfigure[]{\includegraphics[width =0.48\linewidth]{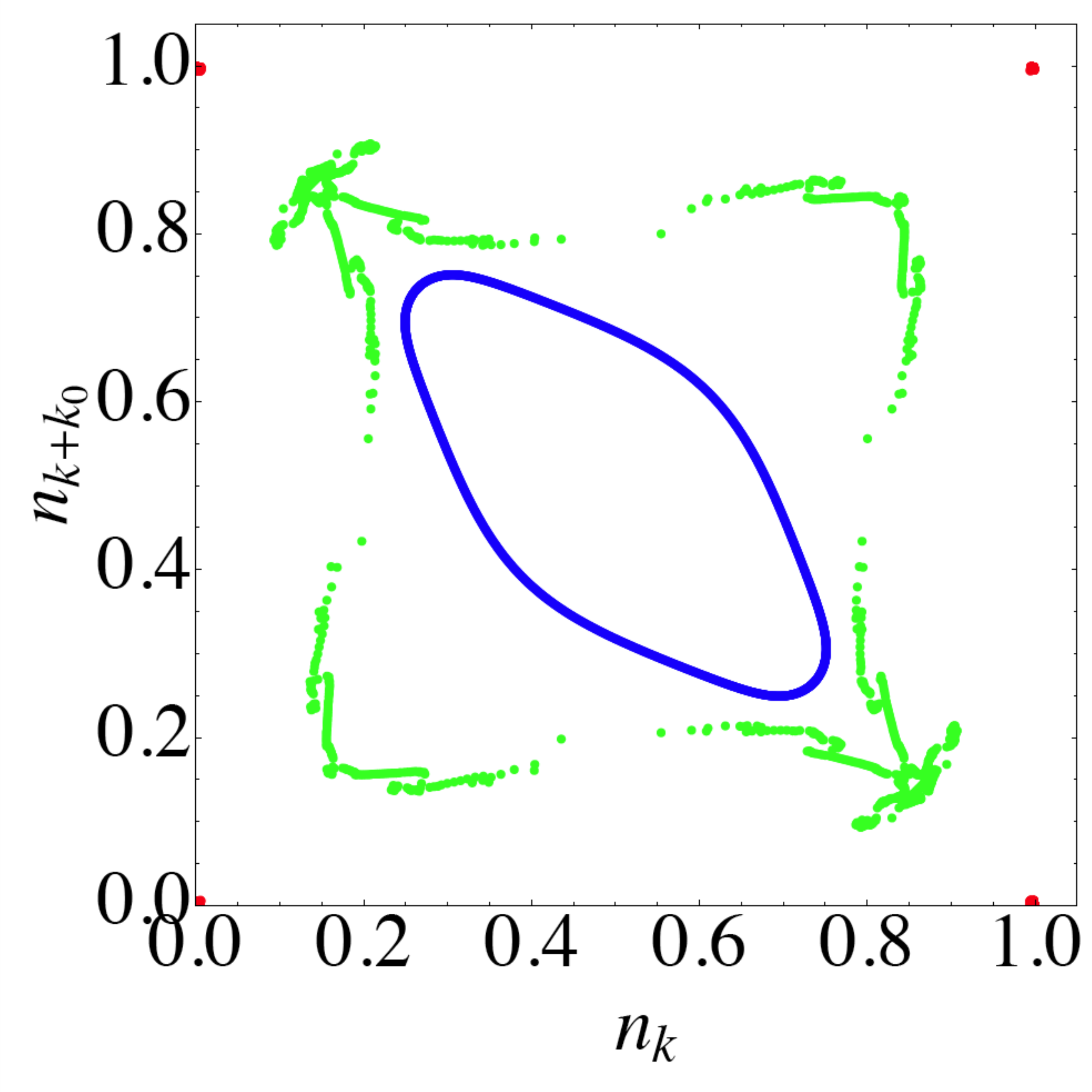}}
 \caption{(color online) Return maps in real space  for (a) and (c) $\lambda=10 (\text{red}),1 (\text{green}),0.5 (\text{blue})$ and the corresponding  return map in momentum space (b)  and (d) $\lambda=0.1 (\text{red}),1 (\text{green}),2 (\text{blue})$. The filling factors are $\nu =0.25$ for the upper panels and $\nu=0.5$ for the lower panels.  The insets are averaged over 50 random phases. }
\label{return}
\end{figure}

Fig.\ref{return} shows the return map
for various rational filling factors $\nu=N_p/N_l=1/4, 1/2$ and different values of $\lambda$.
For $\lambda < 1$, the return maps are smooth curves and correspond to the KAM tori of the extended or metallic phase.
For $\lambda > 1$, the density profile is a discontinuous function, a Cantori. The discreteness of the return map for generic filling factors can be easily understood in the $\lambda\to \infty$ limit, where the wave functions are fully localized and thus the return map can only take the  four possible values $(0,0),(0,1),(1,0),(1,1)$. Exactly at  the transition point $\lambda=1$, the smooth curves become disconnected.

Using the duality transformation, similar return  maps can be drawn in momentum space,
 $n_{k}$ vs
$n_{k\pm k_{0}}$ ($k_{0}=F_{M-1}$). While nearest neighbor sites are connected in position space due to a finite $J$, quasi-momentum components separated by the main reciprocal lattice vectors of the secondary lattice are connected by a finite $V_0$.
 We find  the momentum return maps exhibit an important advantage compared with the local density maps, which is related to the fact that they are insensitive to variations of the phase and retain their pattern when averaged over it. This is not the case in the density maps as shown in the insets of Fig. \ref{return}.

We now consider the case where the filling factor is a ratio of two main Fibonacci numbers  ($\nu=\sigma,1-\sigma$). We refer to those filling factors as irrational filling factors since 
they  approach an irrational
number in the thermodynamical limit. As discussed earlier, this results in a band insulating phase as the Fermi energy $E_F$ lies in the gap,
just outside a filled band.

\begin{figure}[htbp]
\subfigure[]{\includegraphics[width =0.48\linewidth]{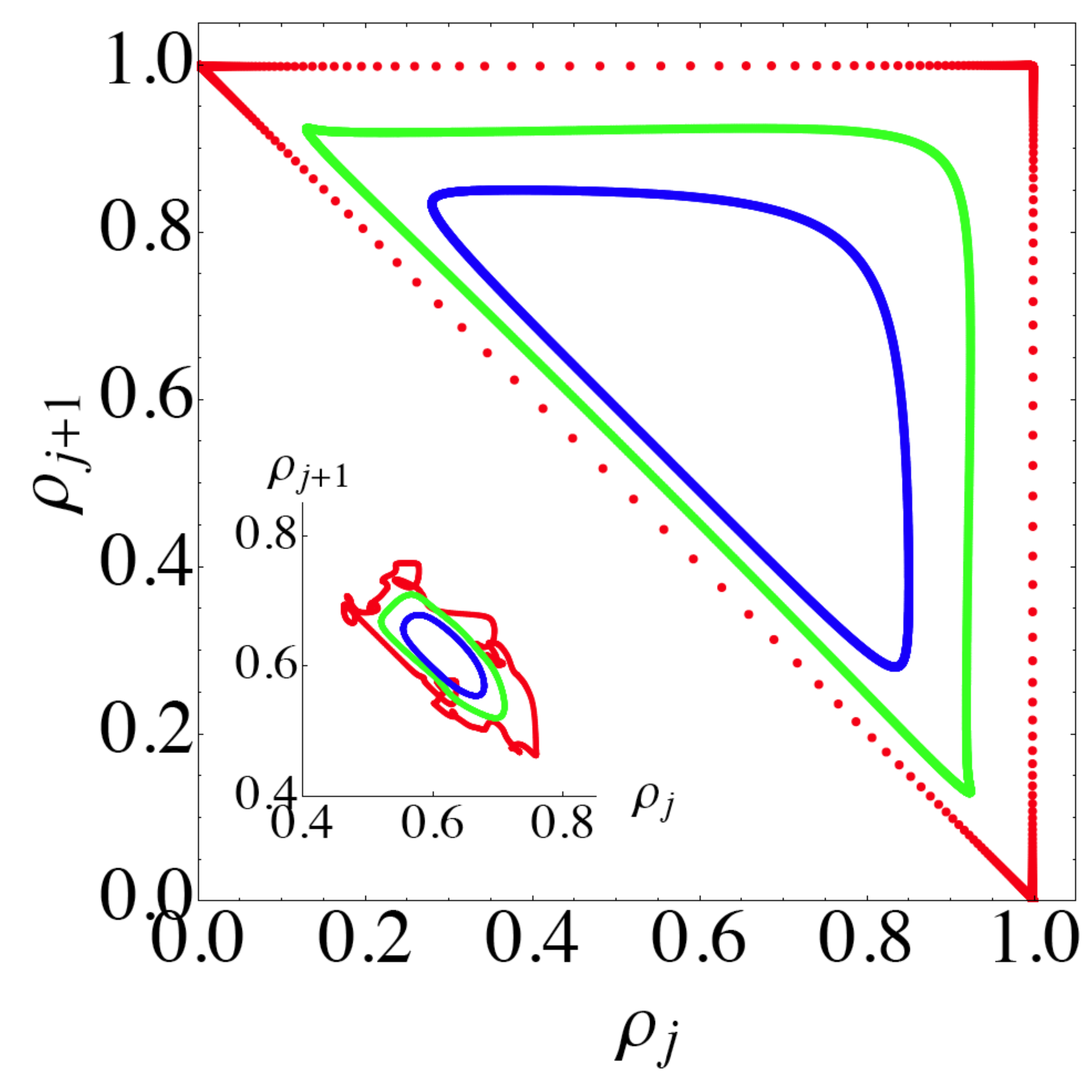}}\quad
\subfigure[]{\includegraphics[width =0.48\linewidth]{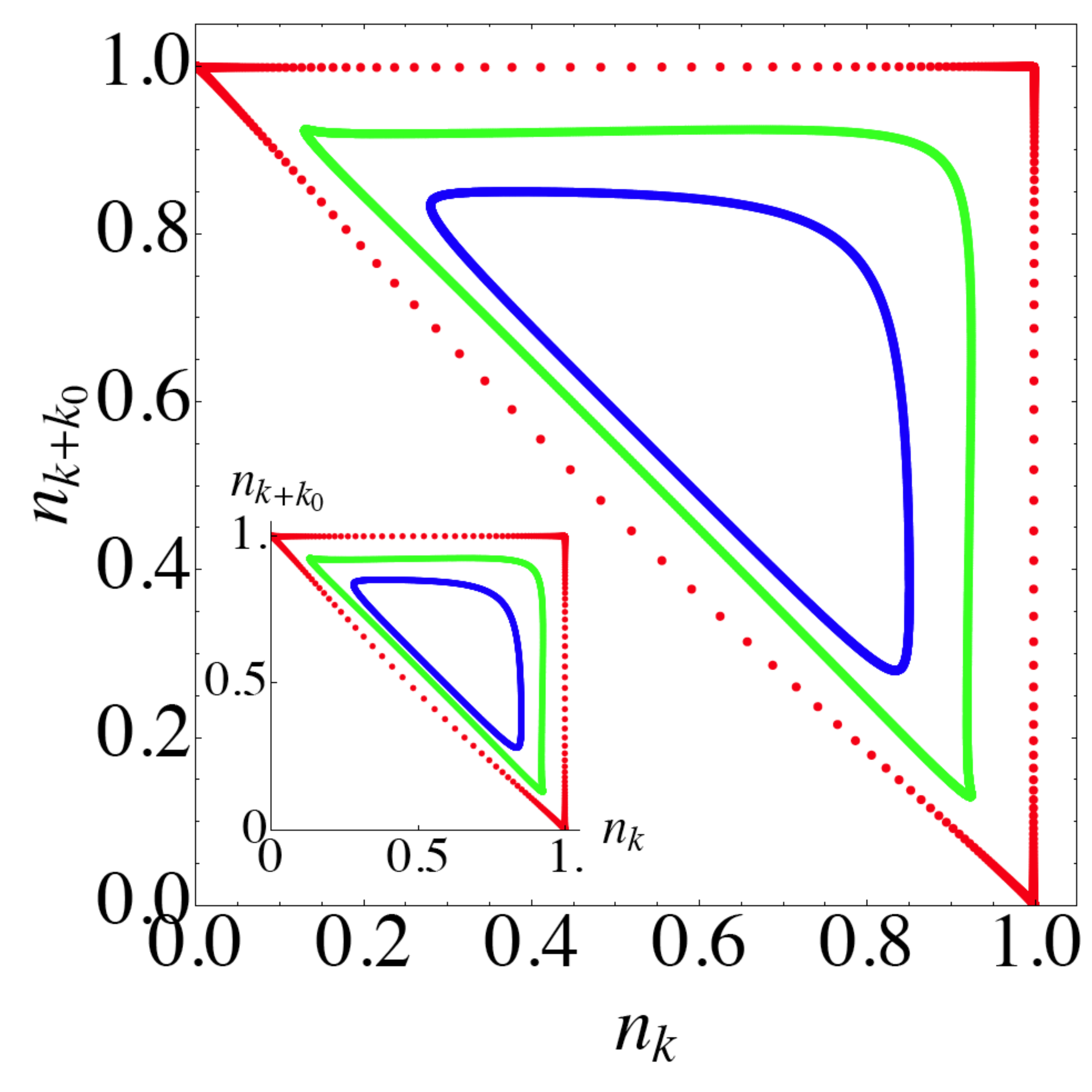}}
\leavevmode
\caption{(color online) Return maps  for the irrational filling factor $\nu=\sigma$. Panel (a) is in real space with $\lambda=10(\text{red}),1(\text{green}),0.5(\text{blue})$ and panel (b) is in momentum space with the dual values $\lambda=0.1(\text{red}),1(\text{green}),2(\text{blue})$.    The insets are averaged over 50 random phases.  }
\label{returnirrational}
 \end{figure}
 
In contrast to the generic or rational filling factors discussed earlier,
the return  map for the irrational filling case is found to remains smooth regardless of the value of $\lambda$
(See Fig.\ref{returnirrational}). This result may appear somewhat counterintuitive, because
as $\lambda \rightarrow \infty$, all the single particle wave functions become localized.
However, an exception to this simple picture occurs near
irrational filling, when the density function exhibits a continuous distribution.
This difference between the rational and the irrational filling is due to the presence
of a group of  paired states centered around the most dominant band edges ( associated with the dominant gaps).  Each of those states is dimerized, by which we mean that they localize at two neighboring sites.
The existence of a pair of such dimerized states is the key to understanding the difference between
rational and the irrational filling: although a single dimerized state causes delocalization near band filling (explaining irrational
filling case), the existence
of a pair can cancel the delocalization effect.
Technical details of this argument are presented in Appendix B, where we show that
the smooth character of the map  at the irrational fillings, $\sigma, \sigma^2$ associated with the leading gaps
can be understood by the breakdown of non-degenerate perturbation theory in the large $\lambda$ limit.

In summary, the band insulating phase
belongs to the  KAM phase irrespective of the strength of the disorder.

\section{Fingerprints of non-linear phenomena in many-body observables}
We now study  the quasi-momentum distribution of polarized fermions, which is directly accessible in the time of flight images of ultracold atoms.
In the following we will show how it imprints a signature of the KAM-Cantori transition and provides experimental realization
of various landmarks of nonlinear systems such as Arnold tongues and bifurcations.
\subsection{Fragmented Fermi Sea}

Fermions in the extended phase have metallic properties. For $\lambda=0$, the single particle eigenstates are fully localized in quasi-momentum space (and thus delocalized in position space) and $n(Q)$ is a
step-like profile: $n(Q)=1$ for $|Q|\leqslant
Q_F$ and $n(Q)=0$ for $|Q|>Q_F$ ($Q_{F}$ is the Fermi momentum).  For  $0<\lambda<1$    the single particle eigenstates localized at $k$ acquire some admixture of other quasi-momentum components ( $n_{k\pm F_{M-1}}$ to leading order in $\lambda$). In this regime  the quasi-momentum distribution retains part of the step-like profile but  gets fragmented into additional structures centered at different reciprocal lattice vectors of the secondary lattice. We call the filled
states centered around $Q = 0$ the main Fermi sea and those around the QP related reciprocal lattice vectors the quasi-Fermi seas.

\begin{figure}
\includegraphics[width =0.8\linewidth,height=0.8\linewidth]{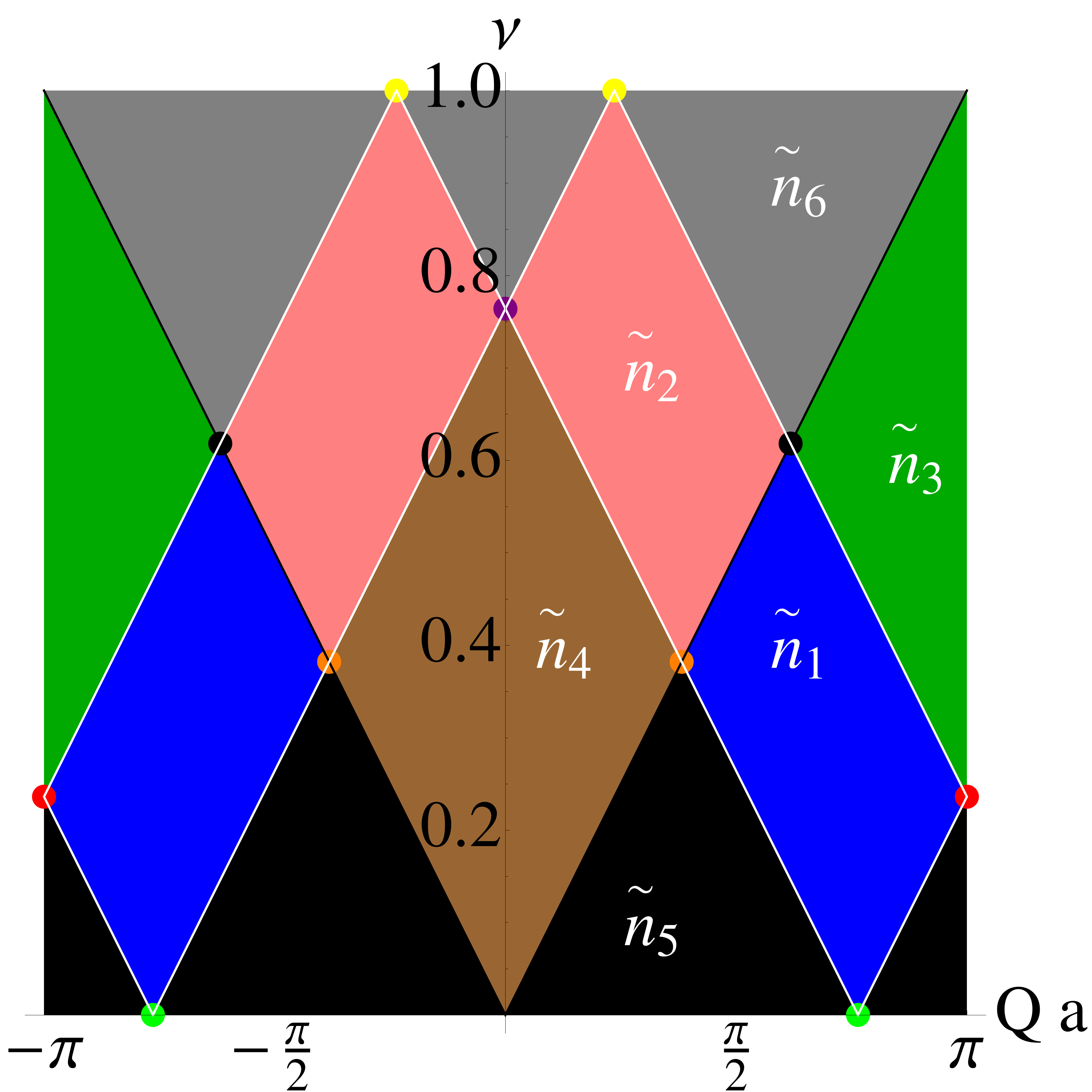}
\caption{(color online) Various boundaries of the fragmented Fermi sea.}
\label{tbound}
\end{figure}

The fragmentation of the quasi-momentum distribution and the development of QP Fermi seas can be understood from first order perturbation theory. To first order in $\lambda$ the momentum landscape becomes fragmented in six regions, shown  in Fig.\ref{tbound} and  given by:

\begin{widetext}

\begin{eqnarray}
\tilde{n}_1&=&\left(\frac{\lambda}{2}\right)^2 \frac{1}{4 \sin^2 \Theta\sin ^2 \left(|Q|a-\Theta\right)} \\
\tilde{n}_2&=&1-\left(\frac{\lambda}{2}\right)^2 \frac{1}{4 \sin^2\Theta \sin ^2 \left(|Q|a+\Theta\right)}\\
\tilde{n}_3&=&\left (\frac{\lambda}{2}\right)^2 \frac{1}{4 \sin^2\Theta}
 \left(\frac{1}{\sin ^2 \left( |Q|a+\Theta\right)}+\frac{1}{\sin ^2 \left(|Q|a- \Theta\right)}\right)\\
\tilde{n}_4&=&1-\left (\frac{\lambda}{2}\right)^2 \frac{1}{4 \sin^2\Theta}
 \left(\frac{1}{\sin ^2 \left( |Q|a+\Theta\right)}+\frac{1}{\sin ^2 \left(|Q|a- \Theta\right)}\right)\\\tilde{n}_5&=&0 \\
\tilde{n}_6&=&1
\end{eqnarray}\label{sixparts} 
Where  $\Theta=\frac{F_{M-2}}{F_{M}}\pi=\frac{|Q_{1}|a}{2}$.
The various regions are delineated by the edges of the   QP Fermi seas, shown by white lines in Fig.\ref{tbound} and  given
  in the $Q\ \text{vs}\ \nu$ space by :
\begin{equation}
\nu=\frac{a}{\pi} |Q(k\pm \sigma N_l)| = \frac{a}{\pi} |Q(k\pm F_{M-1})|
\end{equation}

The  regions $\tilde{n}_{1,2} $, $\tilde{n}_{3,4} $ and $\tilde{n}_ {5,6} $ reflect the characteristic particle-hole symmetry in fermionic systems.
\end{widetext}

\begin{figure}
\subfigure[]{\includegraphics[width=0.91\linewidth]{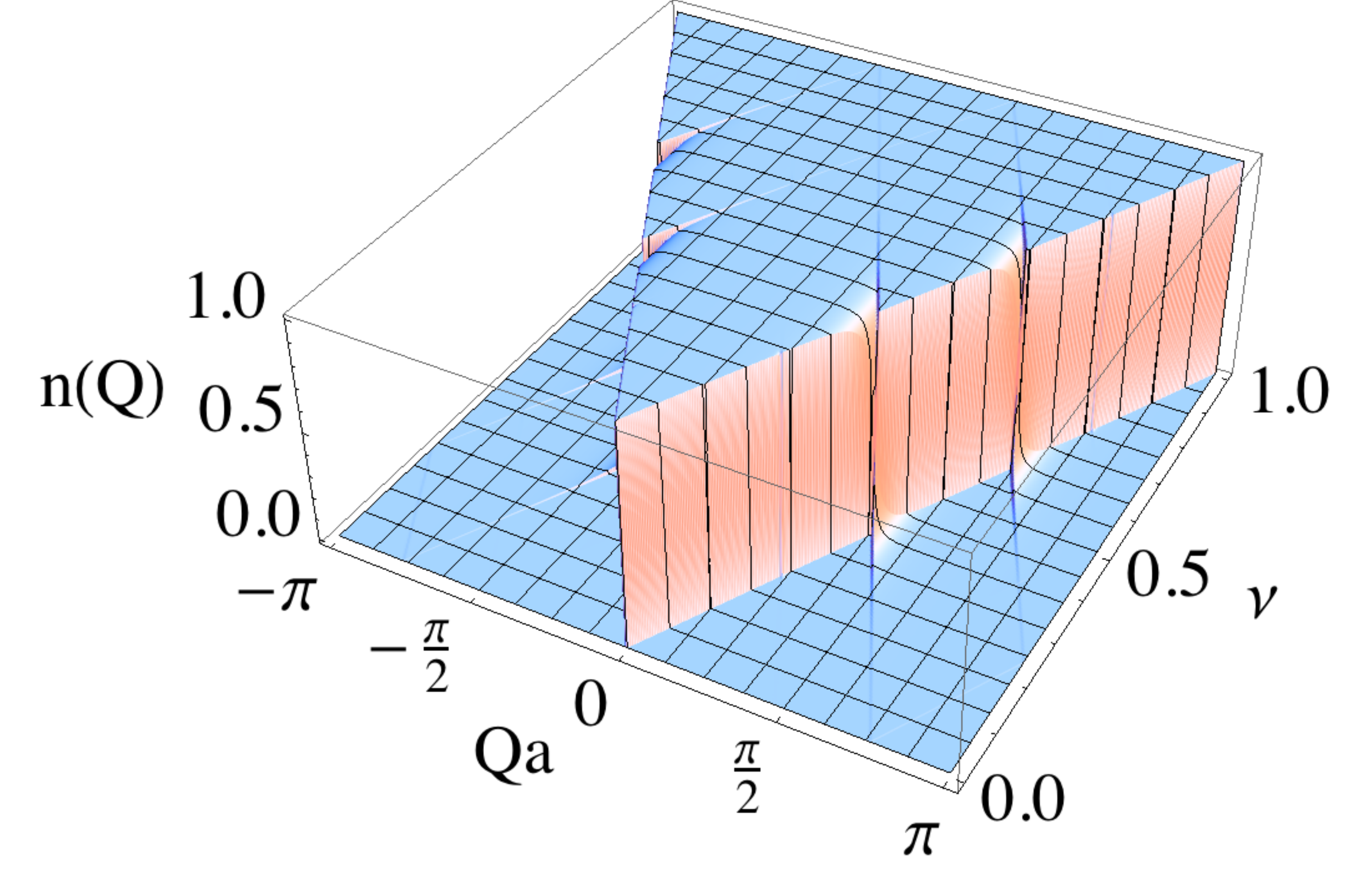}}\\
\subfigure[]{\includegraphics[width=0.91\linewidth]{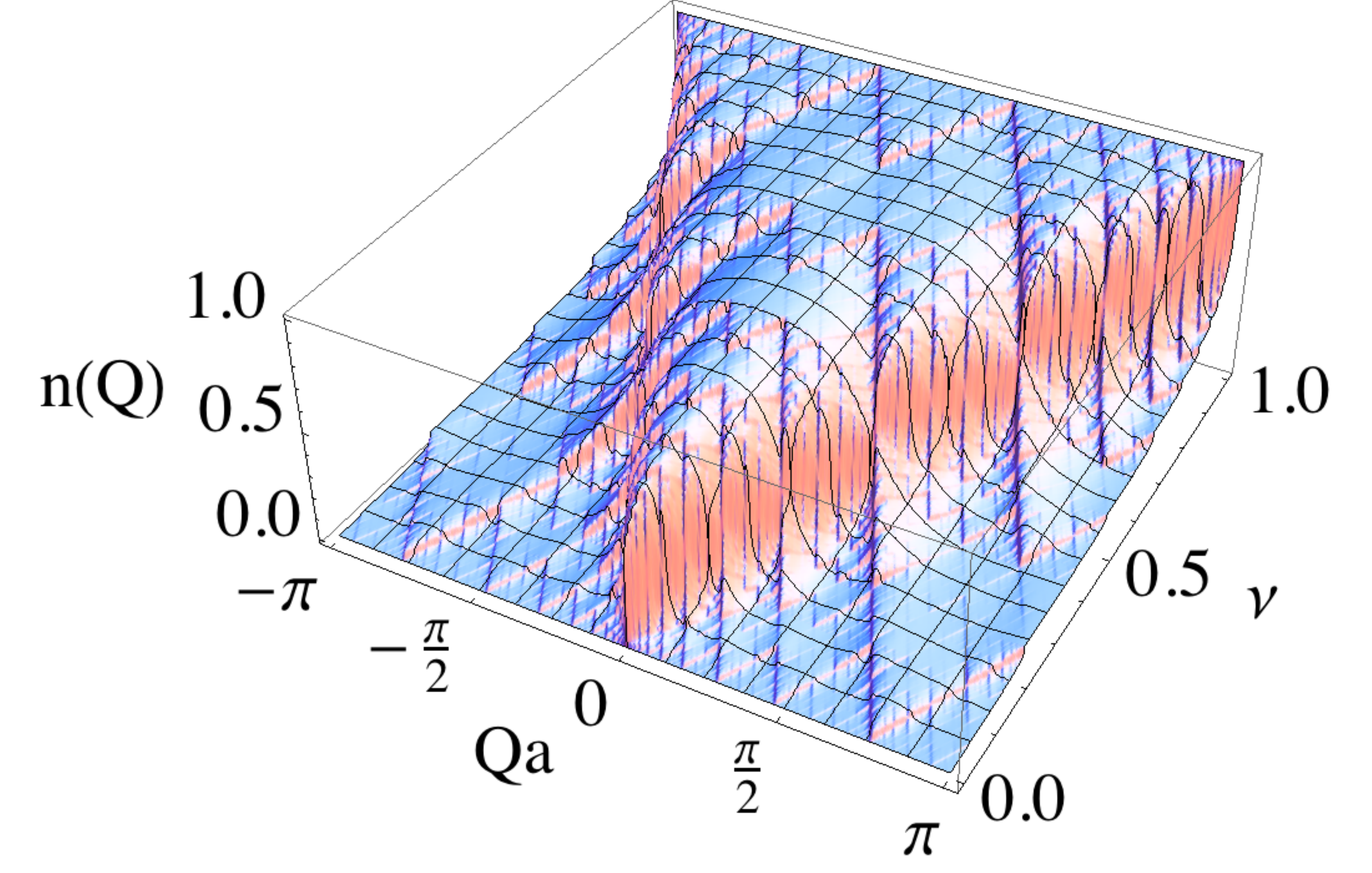}}\\
\subfigure[]{\includegraphics[width=0.91\linewidth ]{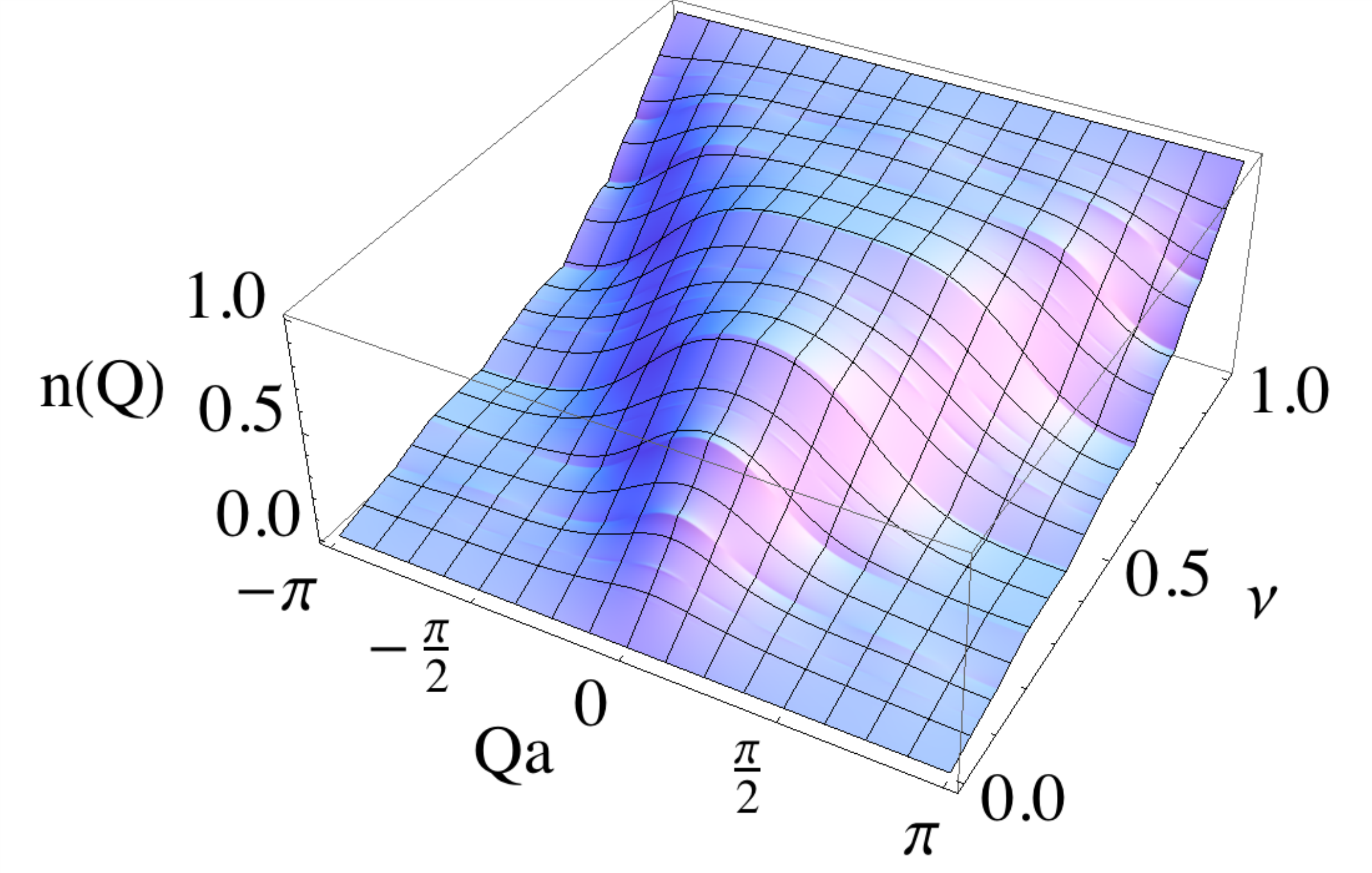}}\\
\caption{(color online) Momentum distribution for (a)$\lambda=0.1$, (b)$\lambda=1$, (c)$\lambda=2$.  All are averaged over 50 random phases to mimic a realistic  experimental case. We can see that the sharp features of these distributions survive phase averaging.}\label{fermisea}
\end{figure}

\begin{figure}
\includegraphics[height =0.61\linewidth]{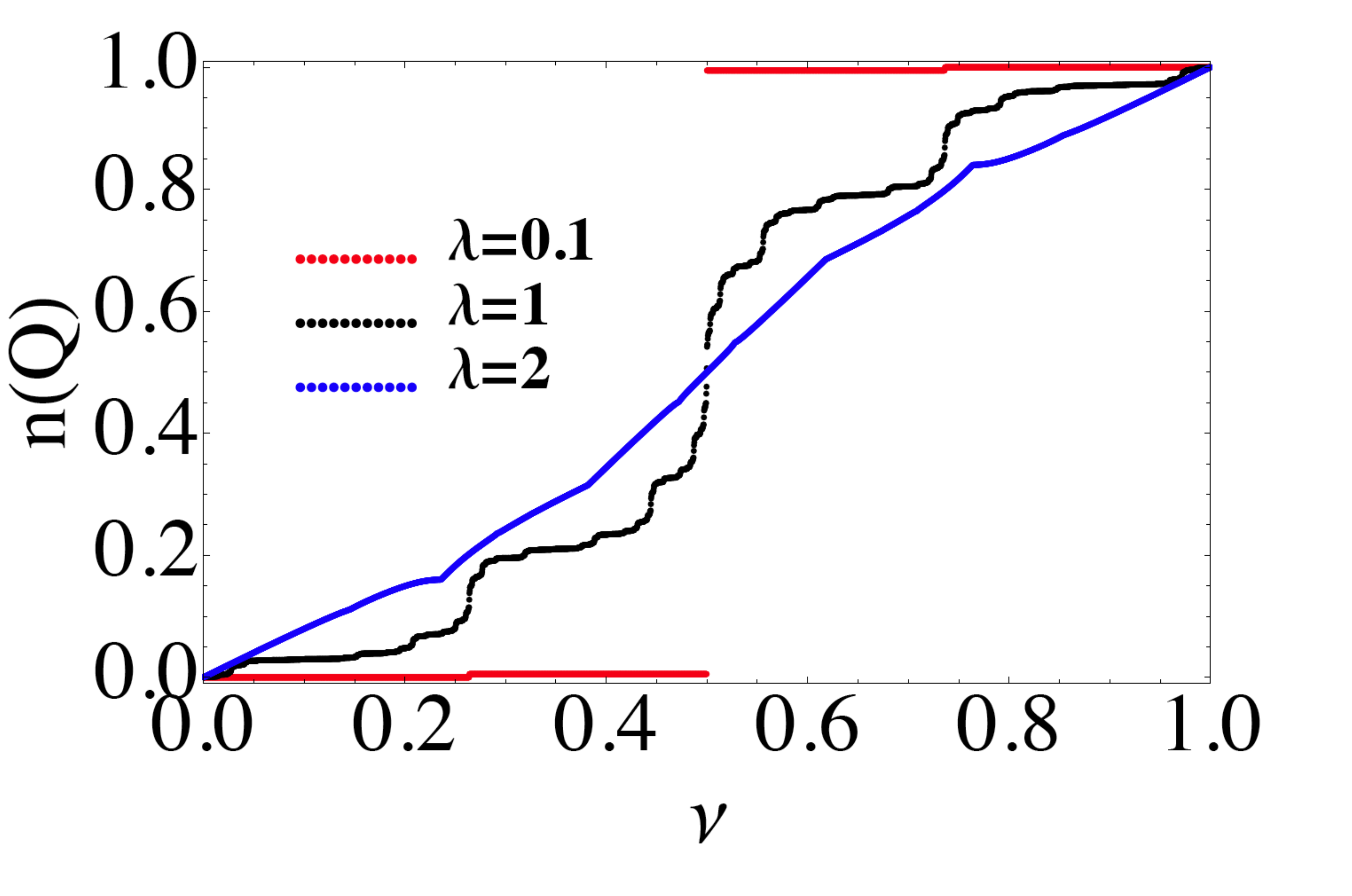}
\caption{(color online) Momentum distribution for fixed $Q(k=1045)$ , $F_{M}=4181$  as a function of filling factors  vs.  the disorder parameter. }\label{nq1}
\end{figure}

\begin{figure}
\subfigure[]{\includegraphics[height =0.61\linewidth]{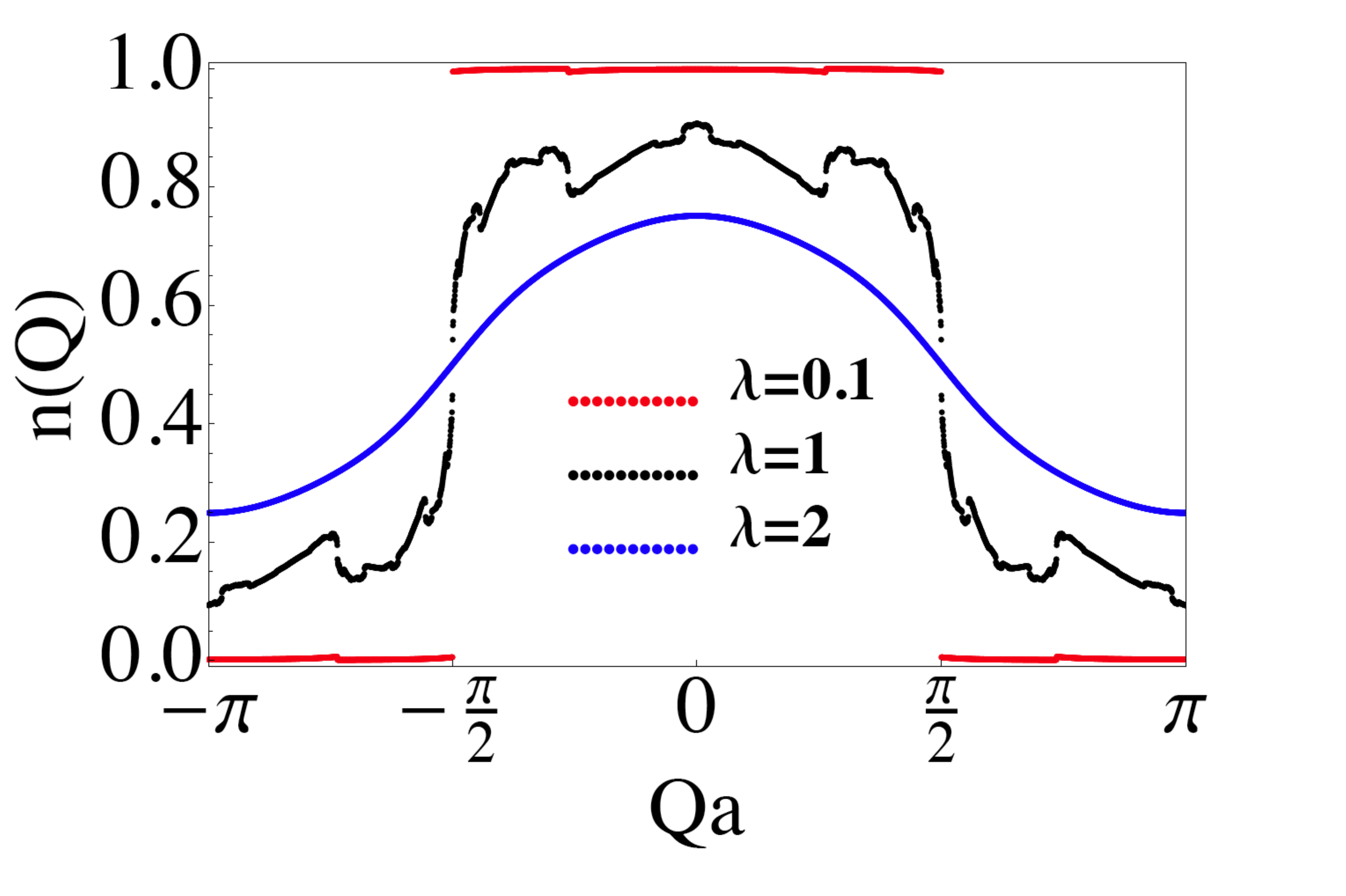}}\quad
\subfigure[]{\includegraphics[height =0.61\linewidth]{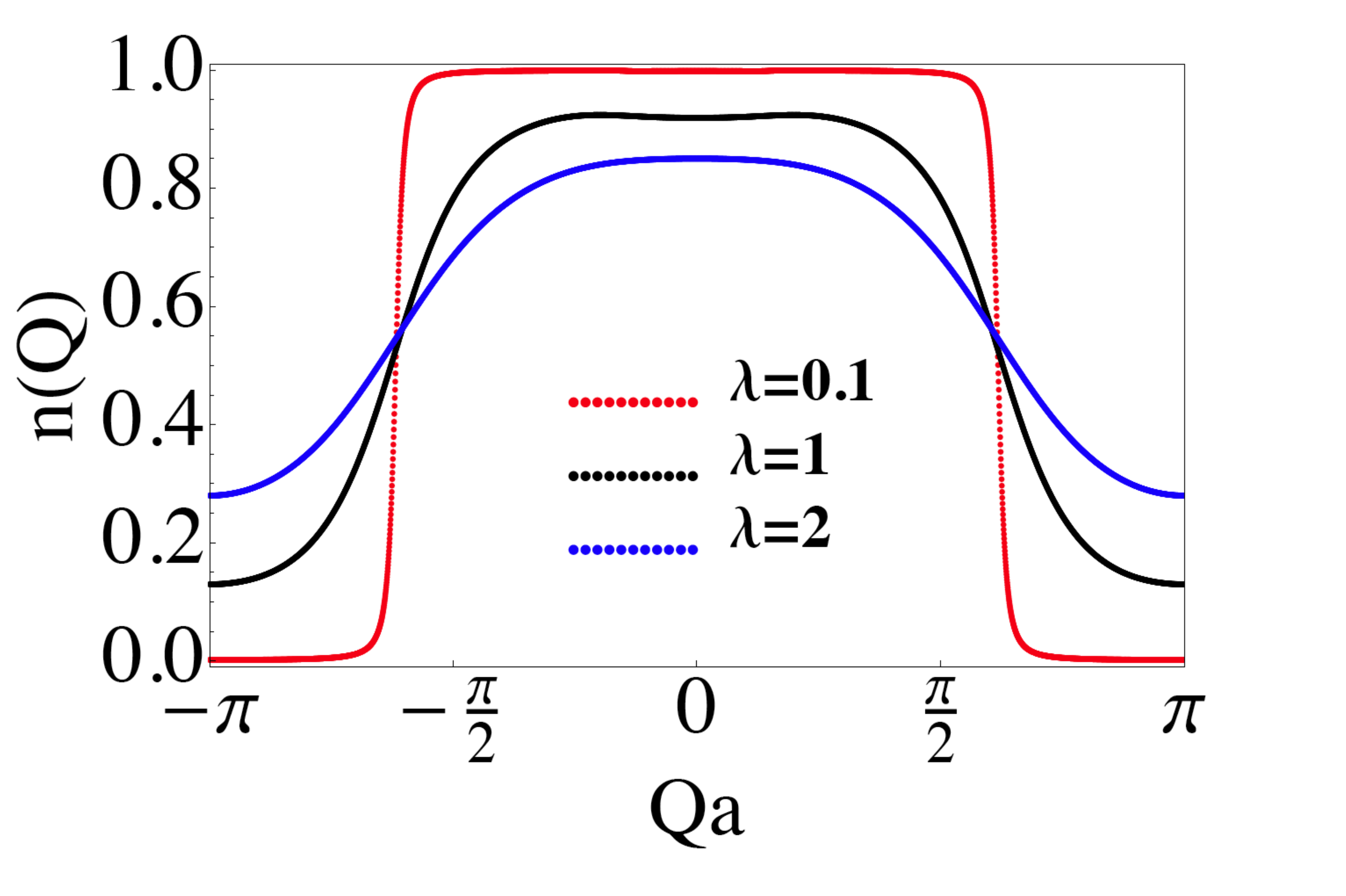}}\\
\leavevmode
\caption{(color online) Momentum distribution vs the disorder parameter. (a) $\nu=0.5$, and   (b) $\nu=F_{M-1}/F_{M}$.}\label{nq2}
\end{figure}

As disorder is increased, more  quasi-Fermi seas become
visible, and with the increase of $\nu$ they overlap in a complicated pattern. Exactly at criticality,  $\lambda=1$, the pattern evolves  into a fractal-like structure. Beyond the critical point,  $\lambda>1$, the fragmented quasi-momentum distribution profile disappears and instead it  becomes a smooth function  of $Q$ and $\nu$.  This behavior  is summarized in Fig. \ref{fermisea} and the  corresponding cross sections for fixed  $Q$ and $\nu$ are displayed in  Fig. \ref{nq1} and Fig. \ref{nq2}.

At irrational filling factors, however, the quasi-momentum distribution profile  remains smooth  regardless of the value of $\lambda$. The smoothness of the momentum distribution at these special filling factors can be understood using the same reasoning as the one used to understand the smooth character of the return map in position space in terms of non-degenerate perturbation theory. In this  case, however,  the states that are coupled are the ones  localized at the  quasi-momentum $Q_{F}+\epsilon_{Q}$ and $-Q_{F}+\epsilon_{Q}$, ( $\epsilon_{Q} a\ll 1$).

The fragmentation of the Fermi sea in the $\lambda< 1 $ regime, and the smooth profile in both the $\lambda>1$ and  band insulator phases  are directly connected, via the self-duality property, to  the discrete nature of the density  return map in the localized phase,  and its smooth  character in the extended and band insulator phases.

An  important point to emphasize again, which is crucial for possible experimental observation of the predicted behavior,  is the  insensitivity of the momentum distribution to variations of the phase $\phi$.  We demonstrate such  insensitivity  by noticing that    the quasi-momentum distributions plotted in Fig.\ref{fermisea} are actually averaged over many different values of $\phi$. Hereafter all  the plots in  momentum space are always averaged over 50 random phases.

\subsection{ Arnold Tongues}
Arnold tongues are mode-locked windows that characterize the periodic dynamics of iconic non-linear systems with competing periodicities.
In the $17$th century Christian Huyghens \cite{Huy} noted that two clocks hanging back to back on a wall tend to synchronize their motion.
In general coupled systems such as coupled pendula or pendula whose lengths vary periodically with time exhibit mode-locking \cite{Arnold}. As  the parameters of a system are varied, it  passes
through regimes that are mode-locked and regimes which are not.

Arnold-like  Tongues also appear in the QP fermionic momentum distribution, where they reflect the complex nature of the physics
induced by  the competing periodicities.

\begin{figure}[hbtp]
\subfigure[]{\includegraphics[width =0.7\linewidth]{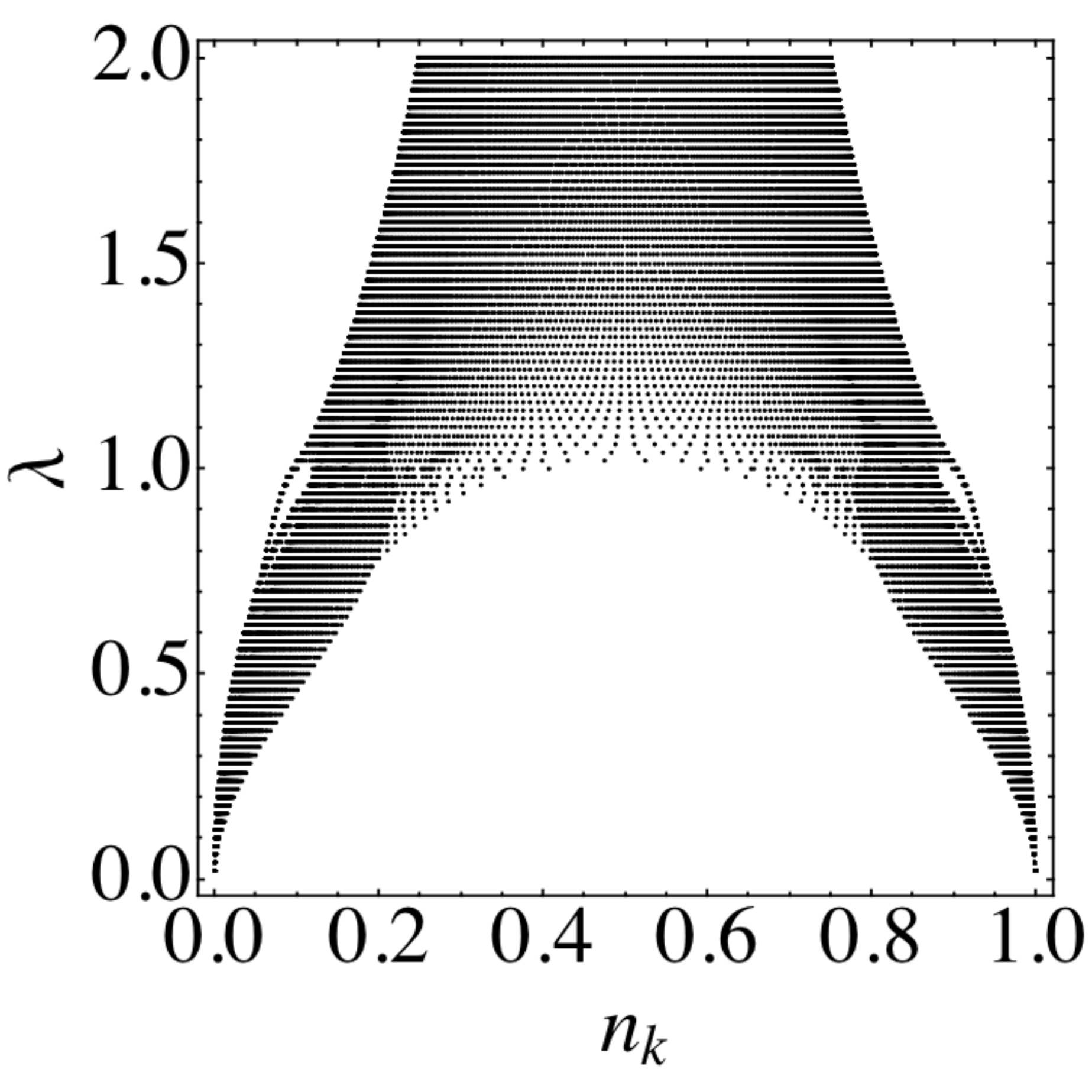}} \quad
\subfigure[]{\includegraphics[width =0.7 \linewidth]{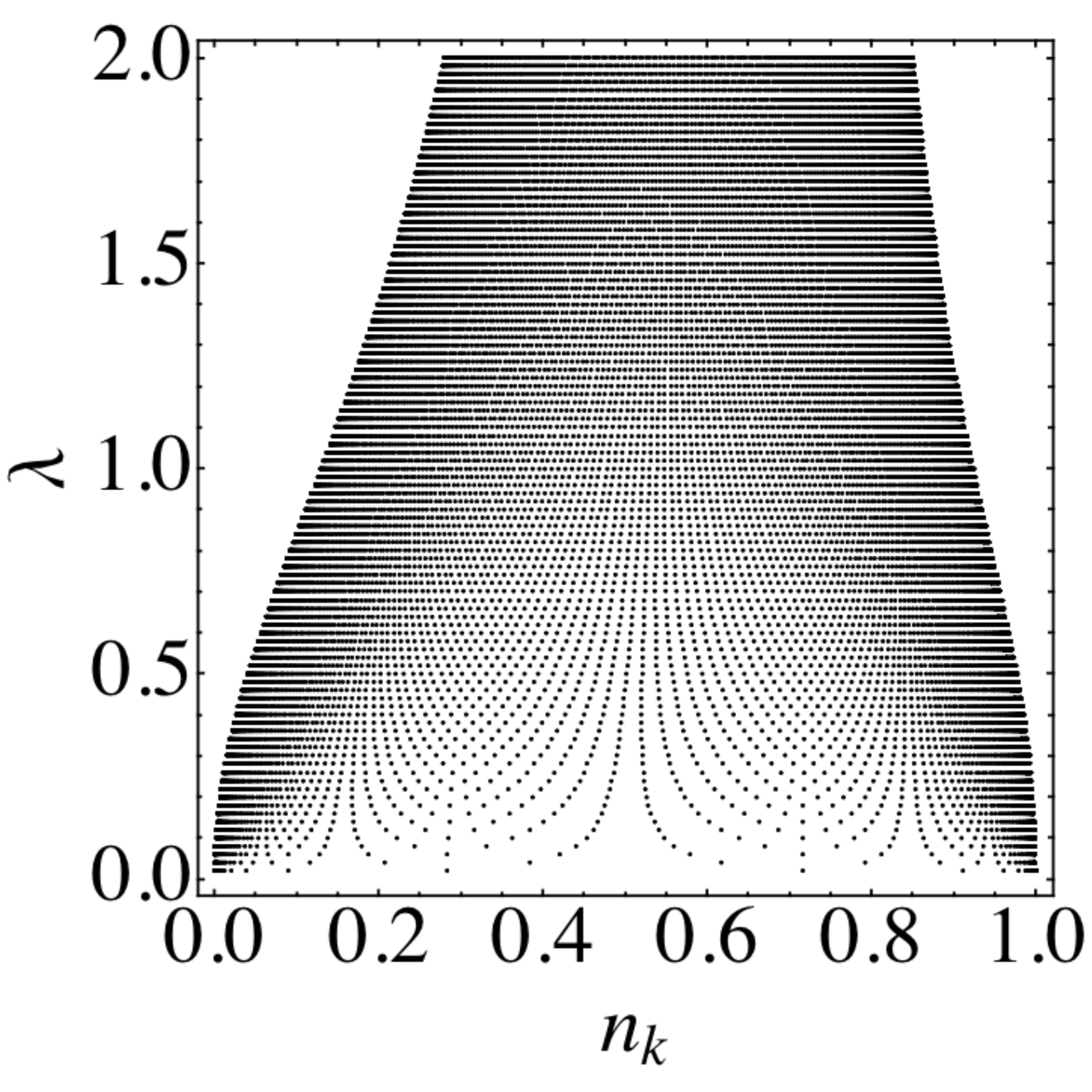}}
\leavevmode
\caption{ Momentum distributions a function of disorder parameter. (a) filling factor  $\nu=0.5$.  Here we see the formation of two Arnold tongues which merge at the critical value $\lambda=1$. (b) $\nu=\sigma$. When $\nu$ is irrational, the system is a band insulator and no Arnold tongues are present. We have checked  the existence of the  similar structure in position space when converting $\lambda$ to $1/\lambda$.} \label{at}
\end{figure}

For $\lambda=0$, the Fermi distribution is a binary distribution with only two values $n(Q)=0, 1$. For finite $\lambda$  and for generic filling factors, two windows  of  values centered around  zero and unity become allowed whose width increases with $\lambda$ as can be derived from a perturbative analysis. We call such windows QP tongues (See Fig. \ref{at}). Exactly at the onset of the  metal-insulator transition the two distributions  overlap, mimicking the Arnold tongue behavior. We have checked that the duality of the Harper equation allows one to observe the formation of analogous structures in position space.
The absence of a metal insulator transition at the irrational filling factors at which the system is a band insulator is also signaled by the disappearance of the Arnold tongues at these fillings, as  shown in Fig.  \ref{at}(b).

\subsection{ Bifurcations}

Bifurcations are common features  observed in   nonlinear dynamical systems,
which occur when a small smooth change made to specific  parameters (the bifurcation parameters) of the  system causes a
sudden `qualitative' or topological change in its behavior. A series of  bifurcations can lead the system from order to chaos.

\begin{figure}[htbp]
\includegraphics[width =0.9\linewidth]{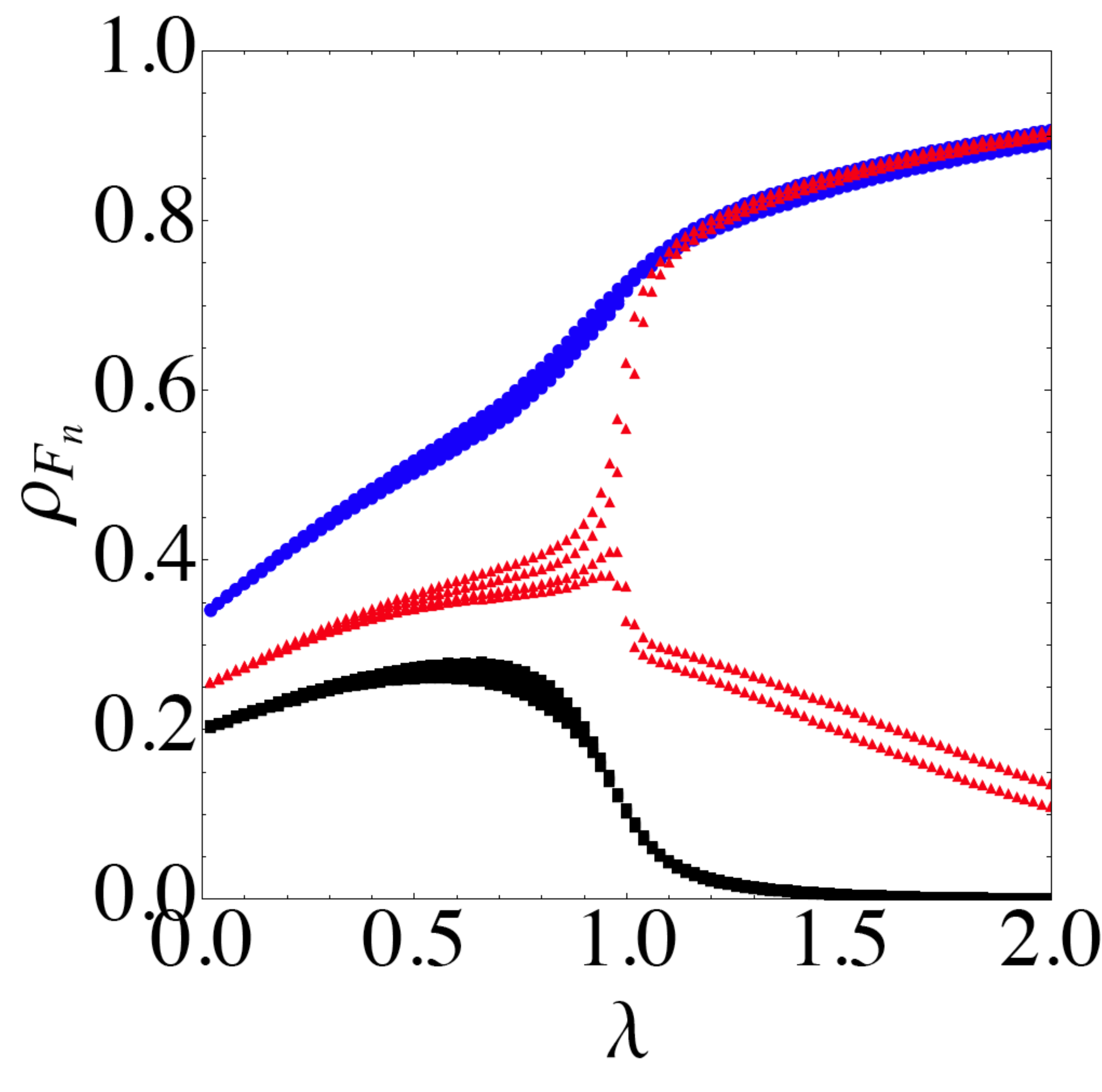}
\leavevmode
\caption{(color online) Bifurcation of local density at Fibonacci sites. 
The central curve (red triangle) corresponds to quarter filling while the lowest (black blocks) and the topmost (blue disks) curves
respectively correspond to $\nu=1/5$ and $\nu=1/3$. Quarter filling is a special case
where the local density at Fibonacci sites is $0$ or $1$ as $\lambda \rightarrow \infty$. For $\nu >0 .25$,
the Fibonacci sites are filled while for $\nu <0.25 $ Fibonacci sites are empty as $\lambda \rightarrow \infty$. We have checked the existence of the same bifurcation phenomena
in the quasi-momentum distribution  but with the
 weak and the strong coupling limits reversed ( $\lambda \to 1/\lambda$). }
\label{bif1}
\end{figure}

The density distribution provides a nice manifestation of a single bifurcation  with $\lambda$ as a bifurcation parameter.
When the density at the various Fibonacci sites is plotted as a function of $\lambda$, for an specific filling factor
which depends on the  value of $\phi$, a bifurcation opens up at $\lambda=1$. In Fig. \ref{bif1} the existence of a bifurcation at quarter filling ($\nu = 1/4$)  when $\phi=3\pi/4$ is shown.

A qualitative understanding  of the bifurcation can be gained   by considering the two limiting regimes, $\lambda\ll 1$ and $\lambda\gg 1$.
 In the weak coupling limit, $\lambda\ll 1$, the local density is uniform and directly proportional  to the filling factor $\nu$.
On the other hand, in the strong coupling limit, a given site remains empty or occupied depending upon whether
the onsite potential is greater than or less than the Fermi energy.

\begin{widetext}
Fibonacci sites $j=F_{n}$ have similar  on-site energies which oscillate  about $\epsilon_c\equiv2\lambda\cos(\phi)$ (assuming $n\approx M$ and $\phi>2\pi F_{M-n}/F_M$ ) as:

\begin{eqnarray}
2\lambda \cos(2\pi \sigma F_n+\phi)=2\lambda\cos \left(2\pi (-1)^{n-1} F_{M-n}/F_M+\phi\right) \left\{
\begin{array}{lr}
<\epsilon_{c}     &\rm\ {\ n \ is \ odd }\\
>\epsilon_{c}     &\rm\ {\ n \ is \ even}
\end{array}
\right. \label{eqfi}
\end{eqnarray}
These oscillations  can be seen in Fig.\ref{mapping} in Appendix A.
\end{widetext} 

Consequently, at  the filling factor, $\nu_c$, at which the Fermi energy matches
$\epsilon_{c}$ (e.g. if  $\phi=3\pi/4$ then $\nu_c\sim 0.25 $) Eq. \ref{eqfi} implies that as  $\lambda$ goes to infinity, even Fibonacci sites ($F_{2l}$) become  empty and   odd  Fibonacci sites ($F_{2l+1}$) become  occupied.
This behavior combined with the monotonic increase of the density with  $\lambda$ in the weak
coupling limit qualitatively explains the observed  bifurcation.

The self-dual behavior described by Eq.\ref{harpereqnk1} implies the existence of the same bifurcation phenomena
in quasi-momentum distribution  but  with the
 weak and the strong coupling limits reversed ( $\lambda \to 1/\lambda$). 

\begin{figure}
\subfigure[]{\includegraphics[width =0.9\linewidth]{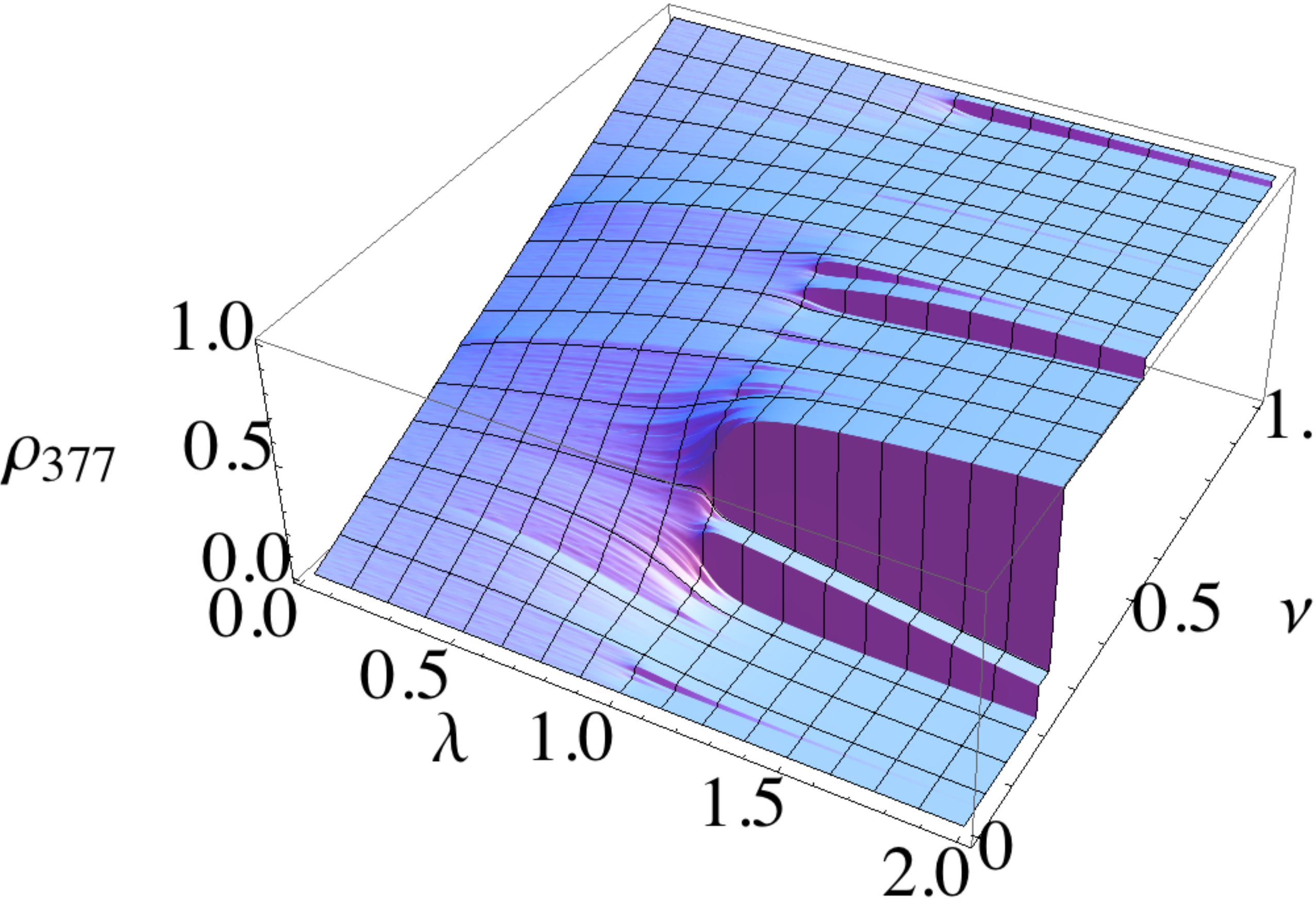}}\quad
\subfigure[]{\includegraphics[width =0.9\linewidth]{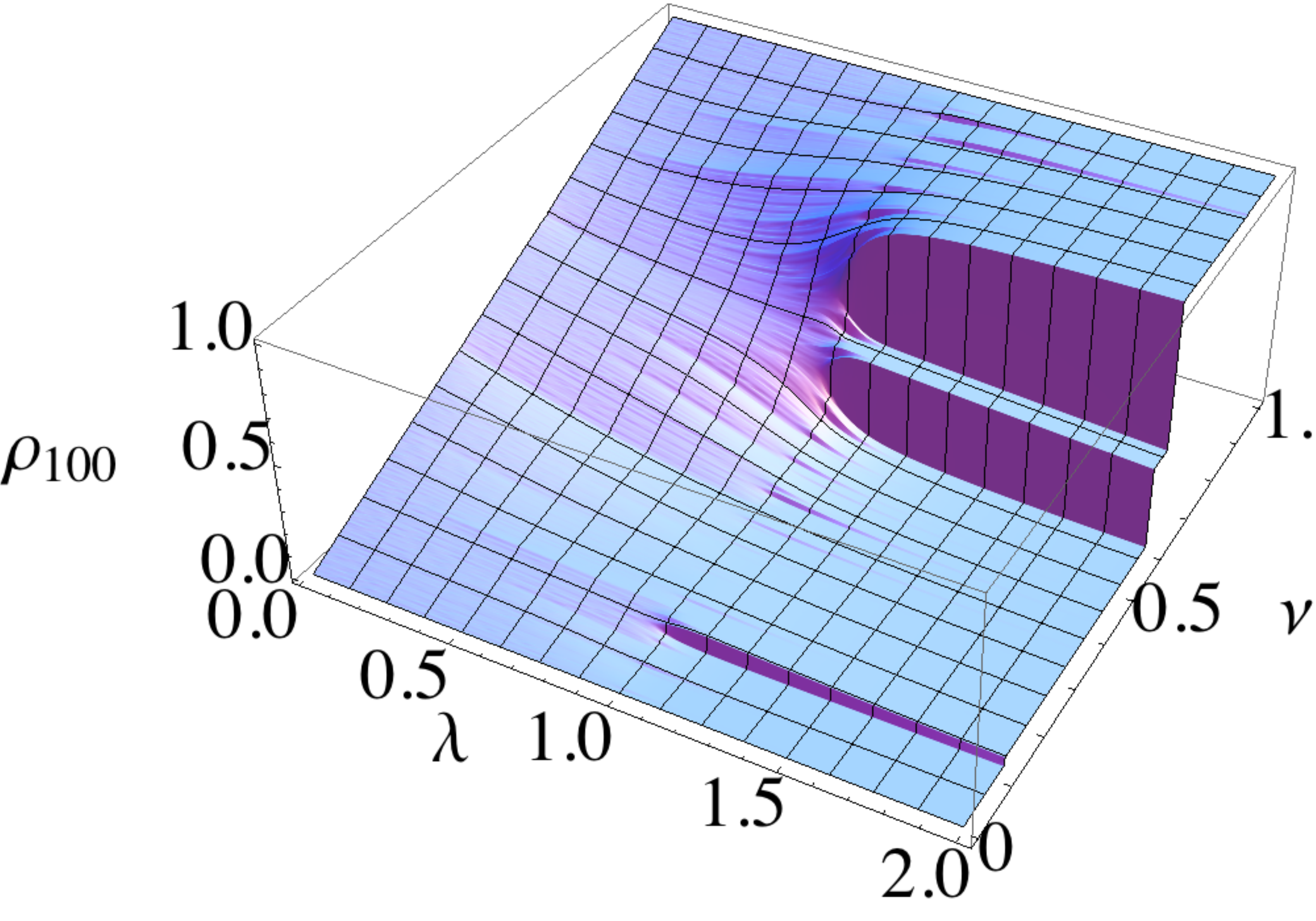}}
\leavevmode
\caption{(color online) Local density distribution  as the disorder parameter and filling factor are varied for (a):  Fibonacci site $j=377$ and (b):  normal site  $j=100$. We have checked the self-dual behavior in the quasi-momentum distribution   with the
 weak and the strong coupling limits reversed ( $\lambda \to 1/\lambda$). }\label{nj}
\end{figure}

When  the local density  is plotted for a generic lattice site as a function of $\lambda$ and $\nu$,
the landscape shows a similar change in topology.
In contrast to  Fig. \ref{bif1} where the filling factor is fixed but different curves corresponding to
different Fibonacci sites  are shown, in Fig. \ref{nj} the density at an specific lattice site is plotted but both the filling factor and disorder strength  are varied.
In this plot besides the large bifurcation there are smaller ones which cannot be explained by studying  the two limiting cases. However, in momentum space they are  qualitatively understood by considering cuts  at a fixed $ \lambda$ as a function of  $\nu$. In these cuts  the  big jump at small $\lambda$  can be identified with the edges of the main Fermi sea in the quasi-momentum profile and  the smaller jumps correspond to edges of the quasi-Fermi seas.

\subsection{ Devil's Staircases}

A Devil's staircase  describes a self similar function $f(x)$, with a hierarchy of jumps or steps. In other words, $f(x)$
exhibits more and more steps as one views the function at smaller and smaller length scale.
The derivative of $f(x)$ vanishes almost everywhere, meaning  there exists a set  of points  of measure 0 such that for all  $x$ outside it  the derivative of $f(x)$ exists and is zero.

Cold atom experiments may provide an opportunity to visualize devil's staircases in the  momentum-momentum
correlations, namely the noise correlations.

  \begin{figure}
\subfigure[]{\includegraphics[width=1.00\linewidth]{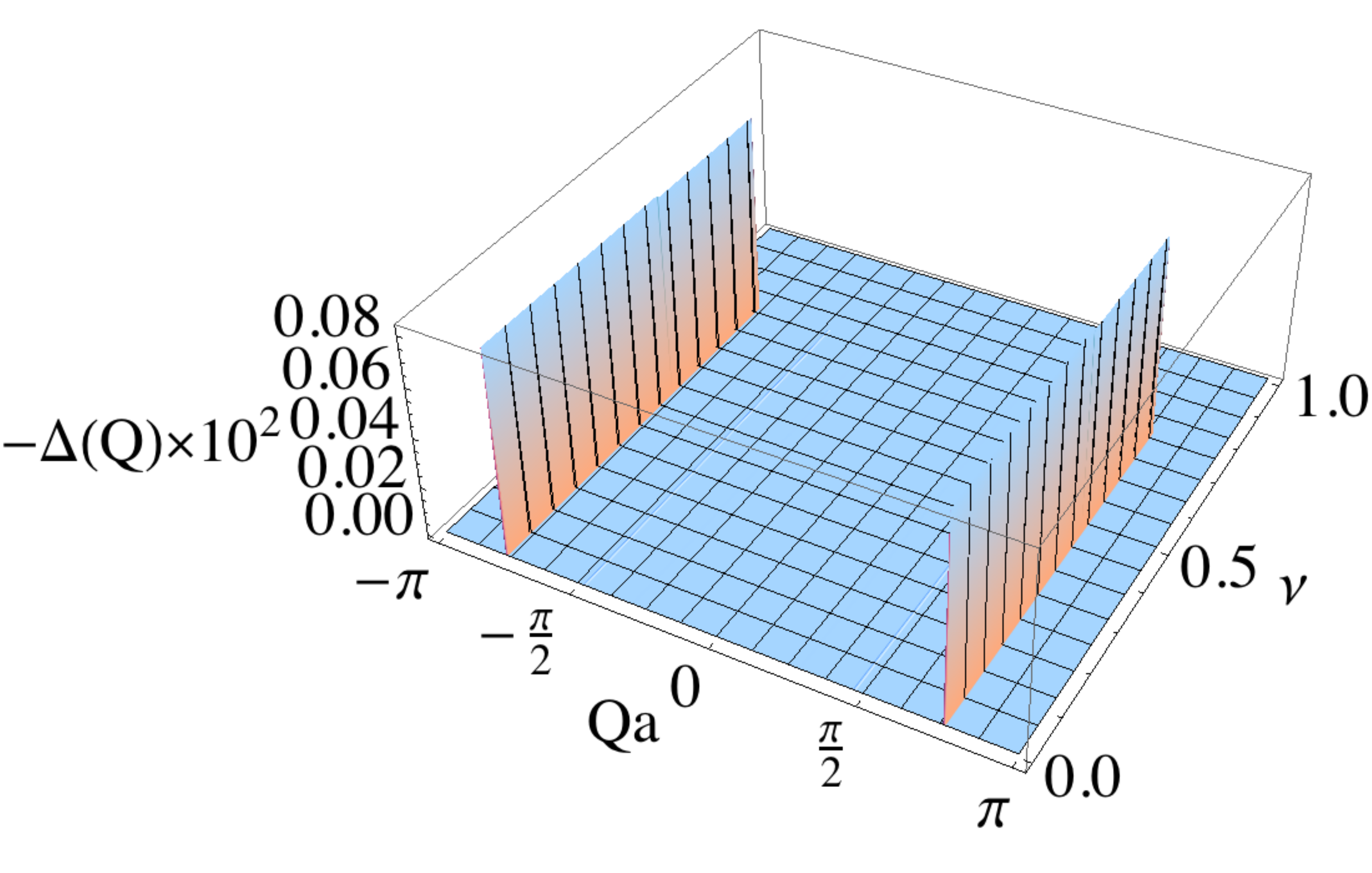}}\\
\subfigure[]{\includegraphics[width=0.91\linewidth]{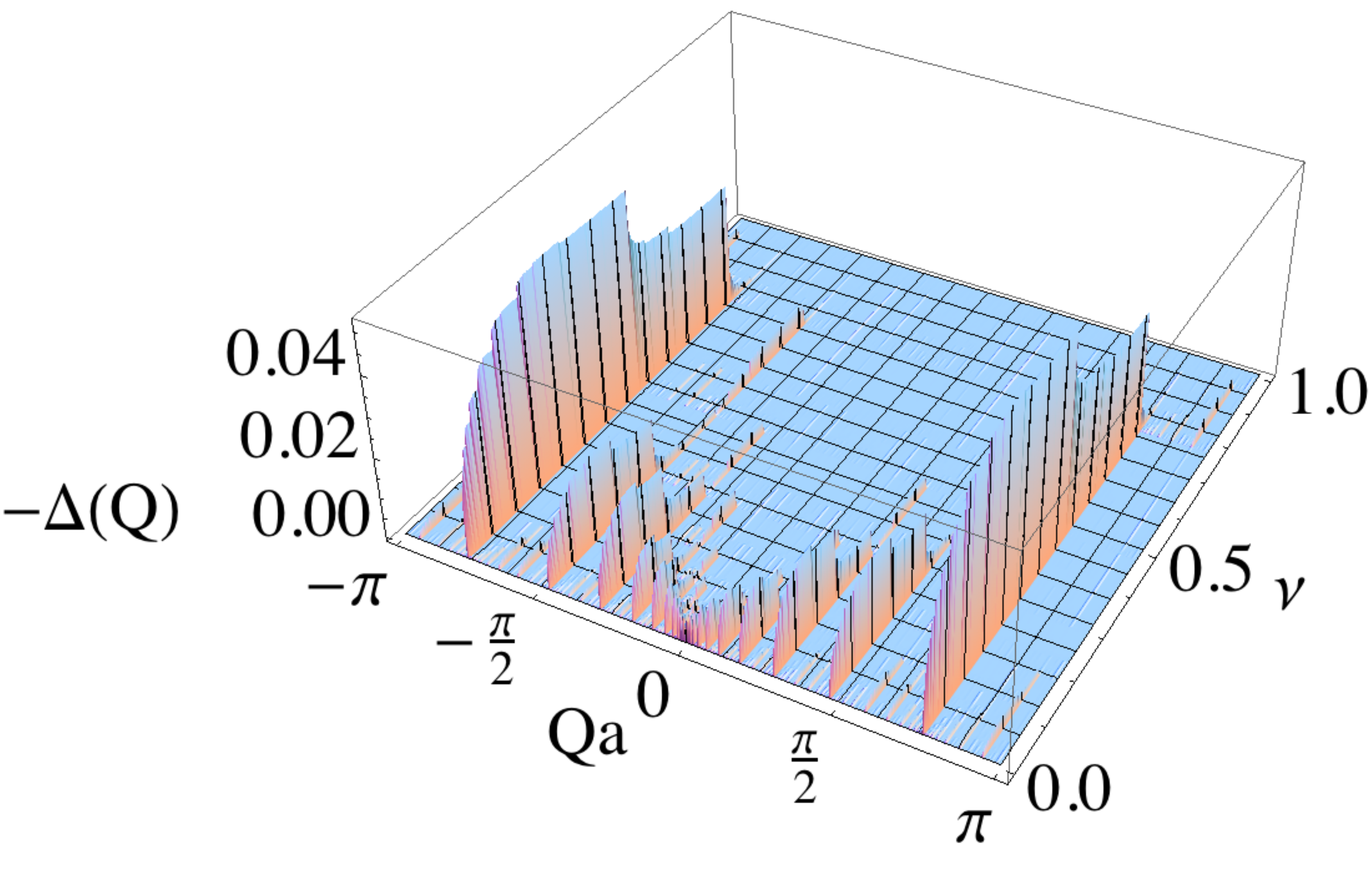}}\\
\subfigure[]{\includegraphics[width=0.91\linewidth]{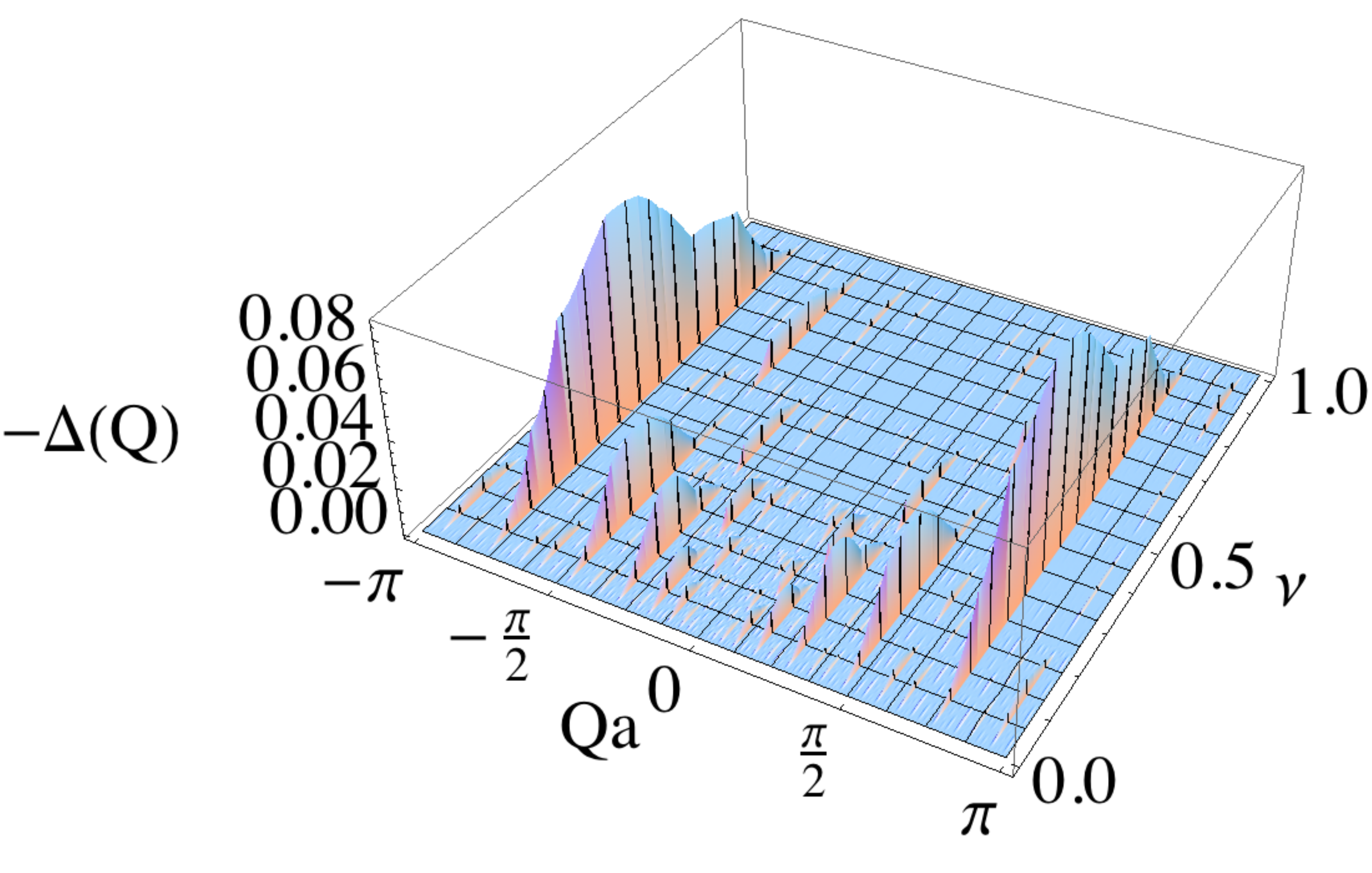}}
\caption{(color online) Noise correlation for (a)$\lambda=0.1$, (b)$\lambda=1$, (c)$\lambda=2$.  All plots display  averaged quantities  over 50 random phases. We can see that structure in the noise correlations survives phase averaging, as for the momentum distributions of Fig. 4.  $\Delta(Q=0)$ is not displayed in those plots.}\label{noisy}
\end{figure}

\begin{figure}
\subfigure[]{\includegraphics[width=0.85\linewidth]{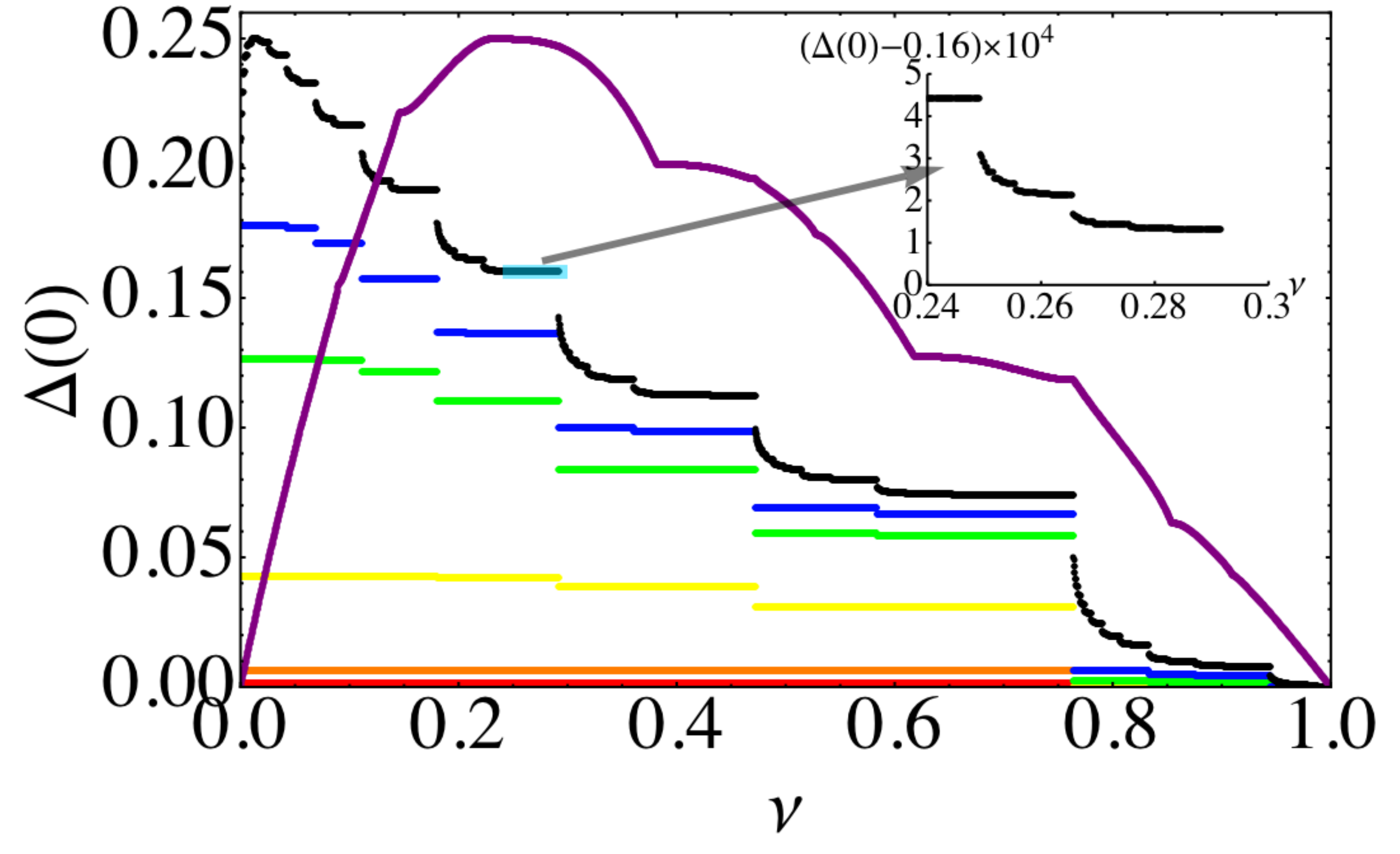}}\\
\subfigure[]{\includegraphics[width=0.85\linewidth]{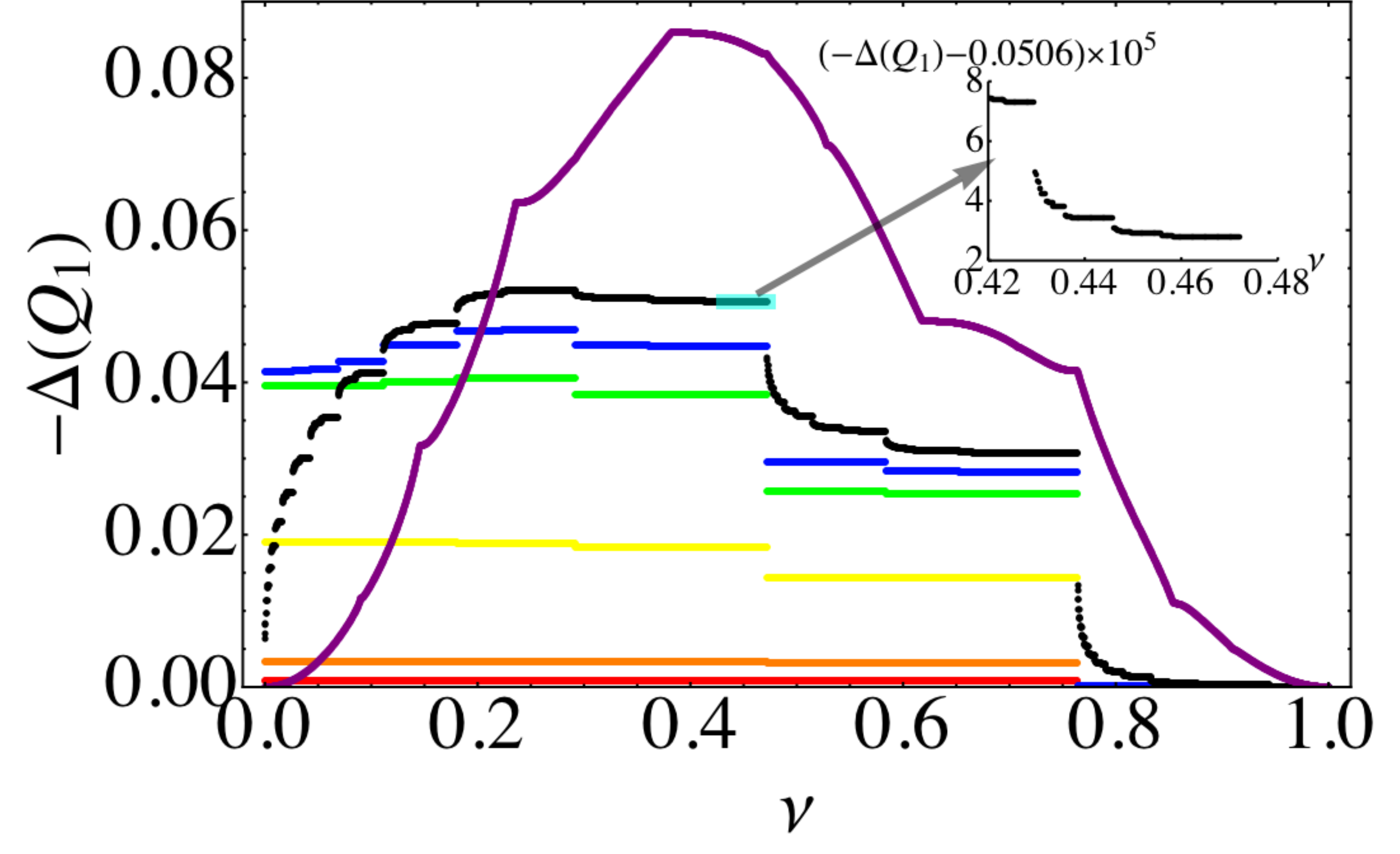}}\\
  \caption{(color online) This figure shows  the evolution of
  $\Delta (Q_{n})$ ( $n=0,1$) as a function of the filling factor for different $\lambda$ values. In the limit  $\lambda\ll 1$, it exhibits  steps occurring at the filling factors $\nu_{ju}^{(m)} (m=1,2,\dots)$ (see text). In the limit  $\lambda\gg 1$,  $\Delta(Q_{n})$ acquires a sinusoidal profile.  The inset shows the self-similar nature of the steps at $\lambda=1$.}
   \label{deltaqn}
\end{figure}

Measurement of noise correlations is an example of
Hanbury-Brown-Twiss interferometry (HBTI),
which is sensitive to intrinsic quantum noise in intensity correlations.
HBTI is emerging as one of the most important tools to provide information beyond that offered by
standard momentum distribution-based characterization of phase coherence. 
The noise correlation pattern in 1D QP bosonic systems has been studied theoretically \cite{Rey2006b, Rey2006d, Rey2007c,Roscilde2008} and has also been
measured experimentally \cite{Guarrera2008}. Here however we focus on the fermionic system.

In the extended phase, noise correlations exhibit a series of plateaus  as $\nu$ is varied, and the number of steps or plateaus increases
as the strength of the disorder $\lambda$ increases.
The origin of this step-like structure with  jumps occurring  at the filling factors $\nu_{ju}^{(n)}=|Q_{n}|a/\pi$, can be
understood from a perturbative argument as follows. For non-interacting fermions, for $Q\neq0$ (See Fig.\ref{noisy}),
$\Delta(Q(k))=-\sum_{m=1}^{N_{p}}\left|\eta_{k}^{(m)}\eta_{0}^{(m)}\right|^{2}$  with $\eta_{k}^{(m)}$ being the
Fourier transform of the $m^{th}$ single-particle eigenfunction.
For $\lambda=0$, the overlap between any two different Fourier components is always zero as only the ground ($m=1$) state has a zero quasi-momentum
component, i.e. $\eta_{0}^{(1)}=1$. For small $\lambda\ll 1$, first order perturbation theory
yields a single step observed at $\nu_{ju}^{(1)}=|Q_{1}|a/\pi$,
as only $\eta_{0}^{(m^{(1)})}$ with $m^{(1)}=\nu_{ju}^{(1)}N_{l}$ is nonzero. 

Quantitatively, the heights of the  steps at $\nu_{ju}^{(1)}$ for $\lambda\ll 1$ are given by:
\begin{eqnarray}
\zeta(0)&=&-2\left(\frac{\lambda}{2}\right)^2\frac{1}{(1-\cos(2\pi \sigma))^2}\nonumber\\
\zeta(Q_1)&=&\left(\frac{\lambda}{2}\right)^2\frac{1}{(1-\cos(2\pi \sigma))^2}
\end{eqnarray}
Where $\zeta(0)$ ($\zeta(Q_{1})$) is the step height for $\Delta(0)$ ($\Delta(Q_{1})$) at filling factor  $\nu_{ju}^{(1)}$. The
minus sign implies a decrease in the noise as the filling factor $\nu$ increases. As $\lambda$ increases, more and more steps are seen
and can be explained using higher order perturbation theory. At criticality (see Figs.\ref{noisy}-\ref{deltaqn}), the steps acquire a hierarchical structure which resembles a devil's staircase and which correlates with the
fractal structure of the energy spectrum  \cite{Ostlund1983,Ostlund1983b}.
In contrast to the momentum distribution,
noise correlations do not show significant differences between the rational and irrational filling factors.

\begin{figure}
\includegraphics[width=1.05\linewidth]{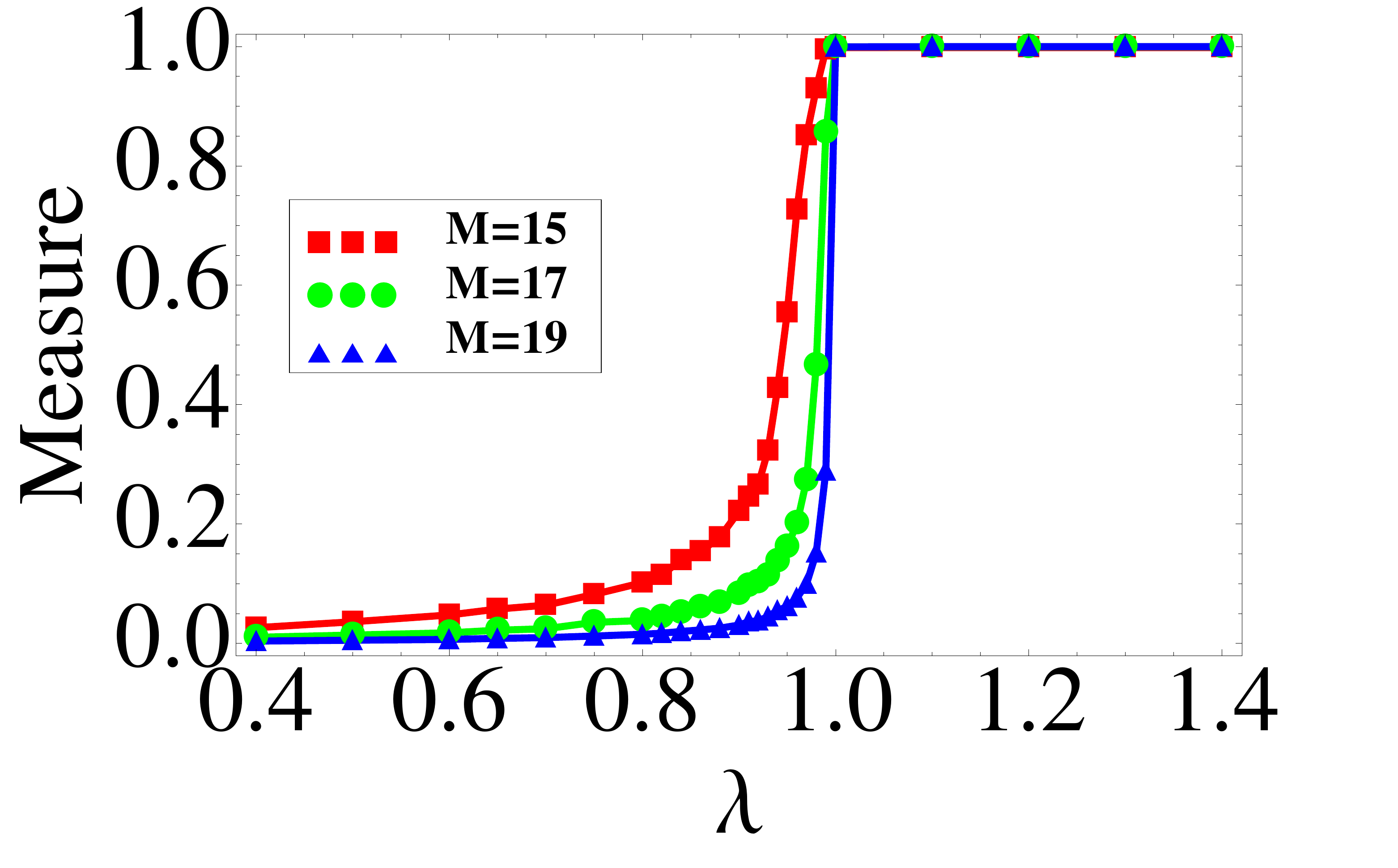}
  \caption{(color online) This figure shows  a measure of the gaps in
  $\Delta (0)$ vs  $\lambda$ for different system sizes, $F_{M}$.  We deem that a gap occurs  if   $|\Delta(0)(N_p+1)-\Delta (0)(N_p)|>\epsilon$   (we choose $\epsilon=10^{-15}$ here).  We can see that with increasing $M$,  the measure of the gaps approaches a step function with the step position at $\lambda=1$. }\label{measure}
\end{figure}

In the localized phase, $\lambda > 1$, we see the smoothing of the step-structure and the noise correlation function  tends towards
a sinusoidal profile $\Delta(Q_{n})\approx \nu \delta_{n,0}-\frac{\sin^{2}(\pi\nu n)}{(\pi n)^{2}}$.

To emphasize the striking difference  between the noise patterns in the extended and localized phases (the step-like vs smooth profile), in Fig. \ref{measure} we plot a measure of  the gaps  developed in $\Delta(0)$ as a function of  $\lambda$.

\section{Summary}

We now summarize our key results:

\begin{enumerate}

       \item { Extended-localized vs KAM-Cantori Phases}
\\

At the many-body level, the connection between the metal-insulator transition
and the KAM-Cantori transition
is signaled in
the  return map of the local density between nearest neighbor sites. For a given angle $\phi$,
the metallic phase with extended single particle wave functions exhibits  a smooth return map.  We identify this with   the preserved invariant KAM tori in phase space at weak perturbations in the corresponding classical system. The localized phase displays a discrete return map, which we identify with the remaining tori or Cantori outside  the perturbative regime. This behavior  is shown in  Fig.\ref{return}.
In addition, for the particular filling factors at which the system becomes a band insulator, the return map remains smooth for any value of $\lambda$ (see Fig. \ref{returnirrational}).

Similar behavior  can be  observed  in  the corresponding momentum distribution return map. However,   while the local density return map loses  these characteristic features  after averaging over $\phi$, the momentum distribution return map remains almost unaffected. The robustness of the momentum return map to phase variations, is ideal for the experimental visualization of the metal-insulator vs KAM-Cantori connection in cold atoms.

\item {Quasi-fractal structures and Arnold Tongues}

The introduction of weak disorder modifies the characteristic  step-function Fermi-sea profile of the quasi-momentum distribution. Additional  step-function structures centered at different reciprocal lattice vectors of the QP lattice appear for $\lambda >0$. We refer to those structures as the ``quasi-Fermi seas".    With increasing  $\lambda$ and $\nu$, the  number and width of the various  quasi-Fermi seas increase  the fragmentation of the momentum distribution, turning it into a complex pattern.  At $\lambda=1$  the fragmentation   becomes maximal and the momentum distribution    evolves into a smooth profile as the system enters the localized phase (See figures \ref{tbound}-\ref{nq2}).
 The overlap of the various
quasi Fermi seas as one approaches  criticality is reminiscent of the Arnold tongues overlap observed in non-linear systems, such  as the circle map, as they enter the chaotic regime \cite{Ott1993}. 

 A more appropriate analogy of such behavior can be observed by considering  the set of  values taken by quasi-momentum distribution   for a  given filling factor.
In the absence of disorder this distribution can only take the values 1 or 0, depending upon
whether the quasi-momentum is greater or smaller than the Fermi quasi-momentum. As the strength of the quasi-periodic lattice increases,  two distributions of   values develop, centered around 0 and 1 respectively. Their width  increases with increasing  disorder  and they  overlap exactly  at criticality (Fig.\ref{at}).

 \item { Bifurcations:} The overlap between Arnold tongues at $\lambda_c$, can be linked, using the space-momentum duality  transformation,  to  the appearance of a bifurcation in  the density profile. The bifurcation occurs at a common but phase dependent filling  for  the various  Fibonacci sites (Fig.\ref{bif1}).
 At generic lattice sites, the filling factor at which the bifurcation  takes place also depends on the lattice site under consideration and can be observed when the local density  is plotted as a function of $\lambda$ and $\nu$ (Fig.\ref{nj}).

\item { Devil's staircases:}  Noise correlations plotted as a function of the filling factor exhibit step-like structures which evolve into a devil's staircase at the onset to the metal-insulator transition ( figures \ref{noisy}-\ref{measure}).

\end{enumerate}

Systems with competing periodicities stand in between periodic and random systems. The richness and complexity underlying such
systems have been studied extensively \cite{Sokoloff1985}. Ultracold atoms  are emerging as a promising candidate to simulate a wide variety
of physical phenomena. Here we have shown they offer opportunities to experimentally realize various paradigms of nonlinear dynamics.

Our focus here was on spin-polarized fermionic systems, since we wanted to look at the simplest consequences of many-body physics in disordered systems. However, Bose-Einstein condensed systems may also be used as tools for laboratory investigation of various predictions made for the quasi-periodic systems based on single-particle
 arguments \cite{Drese1997}.
For example,   it might  be possible
to confirm the  strong coupling universality prediction,
which establishes that  the ratio of the single particle  density at two consecutive Fibonacci sites should be a universal
number \cite{KS}.

{\it Acknowledgments}  A. M. Rey and S. Li acknowledge support from the NSF-PFC grant and NIST.

\appendix

\section{Mapping from position to momentum space from self-duality relationship}
Here we provide an example of the mapping  between  position
coordinates $j$ and  quasi-momentum coordinates $k$  which we use to
link the  quasimomentum-position observables. In the plot we highlight
the Fibonacci sites. A small system size  is used to make the
visualization clearer.

\begin{figure}[htbp]
\includegraphics[width =0.92\linewidth]{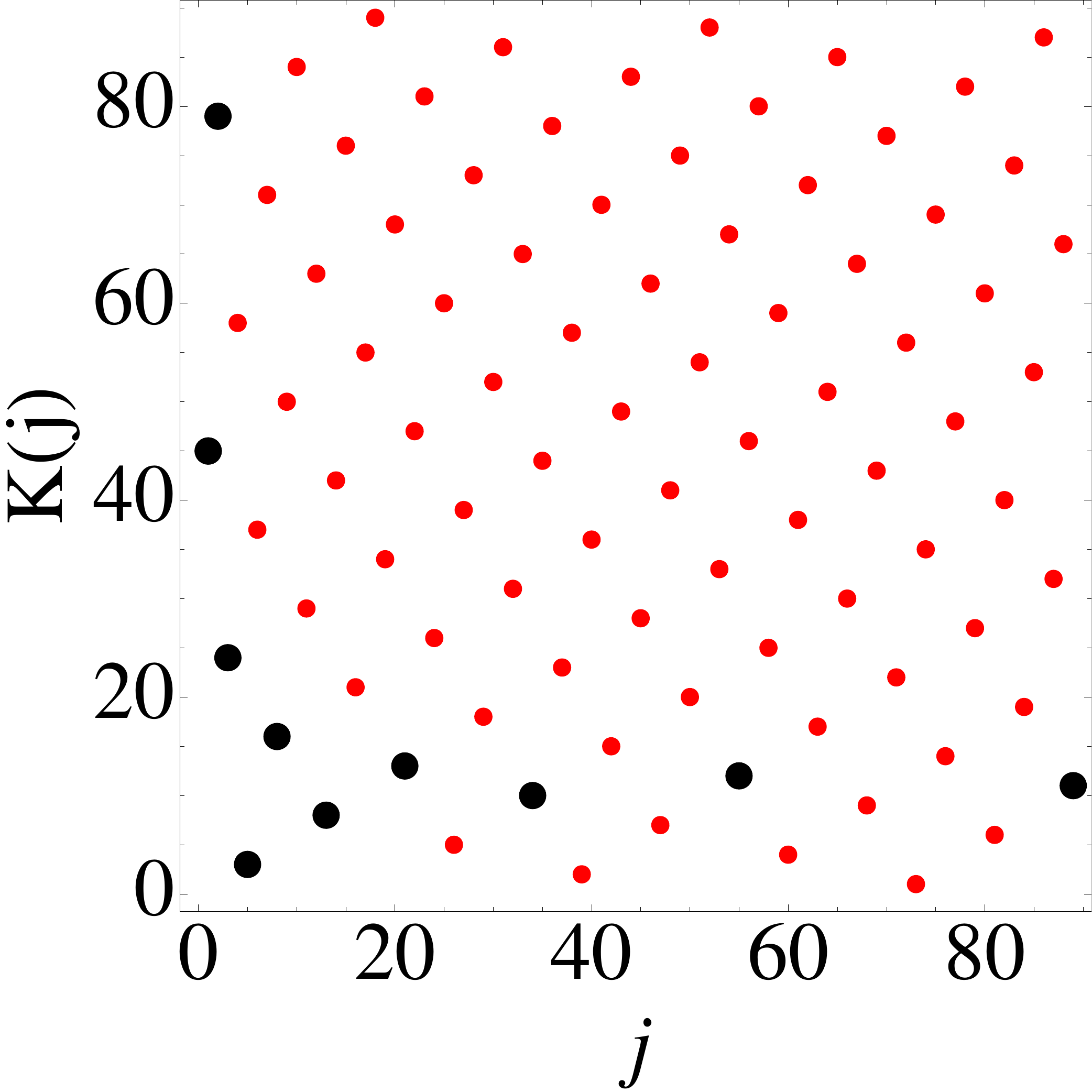}
\caption{(color online)  Mapping from $j$ to $k$ used to determine momentum-space observables from position-space observables and vice versa.  Larger points are Fibonacci sites.  Here we choose $F_{M}=87$, and $K(j)$ is the relation that maps $j$ to $k$.} 
\label{mapping}
\end{figure}
\newpage

\section{Perturbation Theory and Dimerized States}

 We begin our analysis with the Harper equation (\ref{harpereqn}). For $\lambda\rightarrow  \infty$, the single particle wave
functions are localized at individual lattice sites,
$\psi_j^{(m)}=\delta_{j,L_m}$ and $\epsilon^{(m)}=\cos(2\pi \sigma
L_m+\phi)$, where $L_m$ is defined by $\cos(2\pi\sigma
L_{m-1}+\phi)\le \cos(2\pi\sigma L_m+\phi)$. When $ \lambda\gg 1$, we can get the single particle wave function through exact
Diagonalization, or we can use the perturbation theory to obtain the approximate eigenvalues and eigenfunctions.

We focus first  on the site $L_{m}$. Assuming that $2\lambda |\cos(2\pi\sigma L_{m}+\phi)-\cos(2\pi\sigma(L_{m}\pm1)+\phi)| \gg  1$, non-degenerate perturbation theory can be used, and the results are:

\begin{widetext}
\begin{eqnarray}
\epsilon^{(m,0)}&=&2\lambda \cos(2\pi \sigma L_m+\phi)\\
\epsilon^{(m,1)}&=&0\\
\epsilon^{(m,2)}&=& \frac{1}{2\lambda}\bigg(\frac{1}{\cos(2\pi\sigma
L_m+\phi)-\cos(2\pi\sigma (L_m+1)+\phi)}+
\frac{1}{\cos(2\pi\sigma L_m+\phi)-\cos(2\pi\sigma (L_m-1)+\phi)}\bigg) \\
\psi_{j}^{(m,0)}&=&\delta_{L_m,j}\\
\psi_{j}^{(m,1)}&=&-\frac{1}{2\lambda}
\bigg(\frac{\delta_{L_m+1,j}}{\cos(2\pi\sigma L_m+\phi)-\cos(2\pi\sigma
(L_m+1)+\phi)}+\frac{\delta_{L_m-1,j}}{\cos(2\pi\sigma
L_m+\phi)-\cos(2\pi\sigma (L_m-1)+\phi)}\bigg)
\label{pert}
\end{eqnarray}
\end{widetext} where $\epsilon^{(m,0)}$, $\epsilon^{(m,1)}$, $\epsilon^{(m,2)}$ are the zeroth, first and second order terms of the eigenvalue $\epsilon^{(m)}$, with a
similar meaning for  $\psi_j^{(m,0)}$ and
$\psi_j^{(m,1)}$. Also,
$|\psi_j^{(m)}|^2=\delta_{L_m,j}+O((\frac{1}{\lambda})^2)$. So, up to
first order of $\frac{1}{\lambda}$, $\sum_{j}|\psi_j^{(m)}|^2=1$.

 For other sites $L_{m}$ that satisfy $1\approx 2\lambda|\cos(2 \pi \sigma
L_m+\phi)-\cos (2 \pi \sigma (L_m+1)+\phi)|$,   non-degenerate
perturbation theory doesn't work and  degenerate perturbation theory
will be used. In that case, the zero order energies $\epsilon$  and eigenfunctions $\psi$ satisfy:

\begin{widetext}
\begin{eqnarray}
\left(
\begin{array}{ll}
 2 \lambda   \cos (2 \pi \sigma L_m+\phi ) & \ \ \ \ -1   \\
   -1   & 2 \lambda  \cos (2\pi\sigma(L_m+1)+\phi)
\end{array}
\right)
\left(
\begin{array}{l}
\psi_{L_m}^{(m,0)}\\
\psi_{L_m+1}^{(m,0)}
\end{array}
\right) =\epsilon \left(
\begin{array}{l}
\psi_{L_m}^{(m,0)}\\
\psi_{L_m+1}^{(m,0)}
\end{array}
\right)
\label {degen}
\end{eqnarray}

The above equations result in a pair of energies which we denote as $\epsilon_u$ and $\epsilon_d$:
\begin{eqnarray}
\epsilon_d&=&\lambda\{\cos(2\pi \sigma L_m+\phi)+\cos(2\pi\sigma(L_m+1)+\phi)\}-\sqrt{1+\Lambda_m^2}\\
\epsilon_u&=&\lambda\{\cos(2\pi \sigma L_m+\phi)+\cos(2\pi\sigma(L_m+1)+\phi)\}+\sqrt{1+\Lambda_m^2}
\end{eqnarray}
\end{widetext} where $\Lambda_m=\lambda(\cos(2\pi \sigma L_m+\phi)-\cos(2\pi\sigma(L_m+1)+\phi))$. The corresponding orthogonal eigenfunctions
can be written as:
\begin{eqnarray}\psi_d=\left(\begin{array}{l}A\\B\end{array}\right) \quad \rm and\quad  \psi_u= \left(\begin{array}{l}B\\-A\end{array}\right)
\end{eqnarray}
where $A=\sqrt{\frac{\sqrt{\Lambda_m^2+1}-\Lambda_m}{2\sqrt{\Lambda_m^2+1}}}$, $B=\sqrt{\frac{\sqrt{\Lambda_m^2+1}+\Lambda_m}{2\sqrt{\Lambda_m^2+1}}}$. We can choose $\alpha$ ($\beta$) to be the greater (lesser) of $A$ and $B$:  $|\alpha|^2=\frac{\sqrt{1+\Lambda_{m}^2}+|\Lambda_{m}|}{2\sqrt{1+\Lambda_{m}^2}}$,  $|\beta|^2=\frac{\sqrt{1+\Lambda_{m}^2}-|\Lambda_{m}|}{2\sqrt{1+\Lambda_{m}^2}}$.
From these results one can see that  $|\alpha|^{2}$ increases as $|\Lambda_{m}|$ increases. For example,  when $|\Lambda_{m}|=0$, $|\alpha|^{2}=0.5$ and  when  $|\Lambda_{m}|=\infty$, $|\alpha|^{2}=1$.

\begin{figure}
\subfigure[]{\includegraphics[width=1\linewidth]{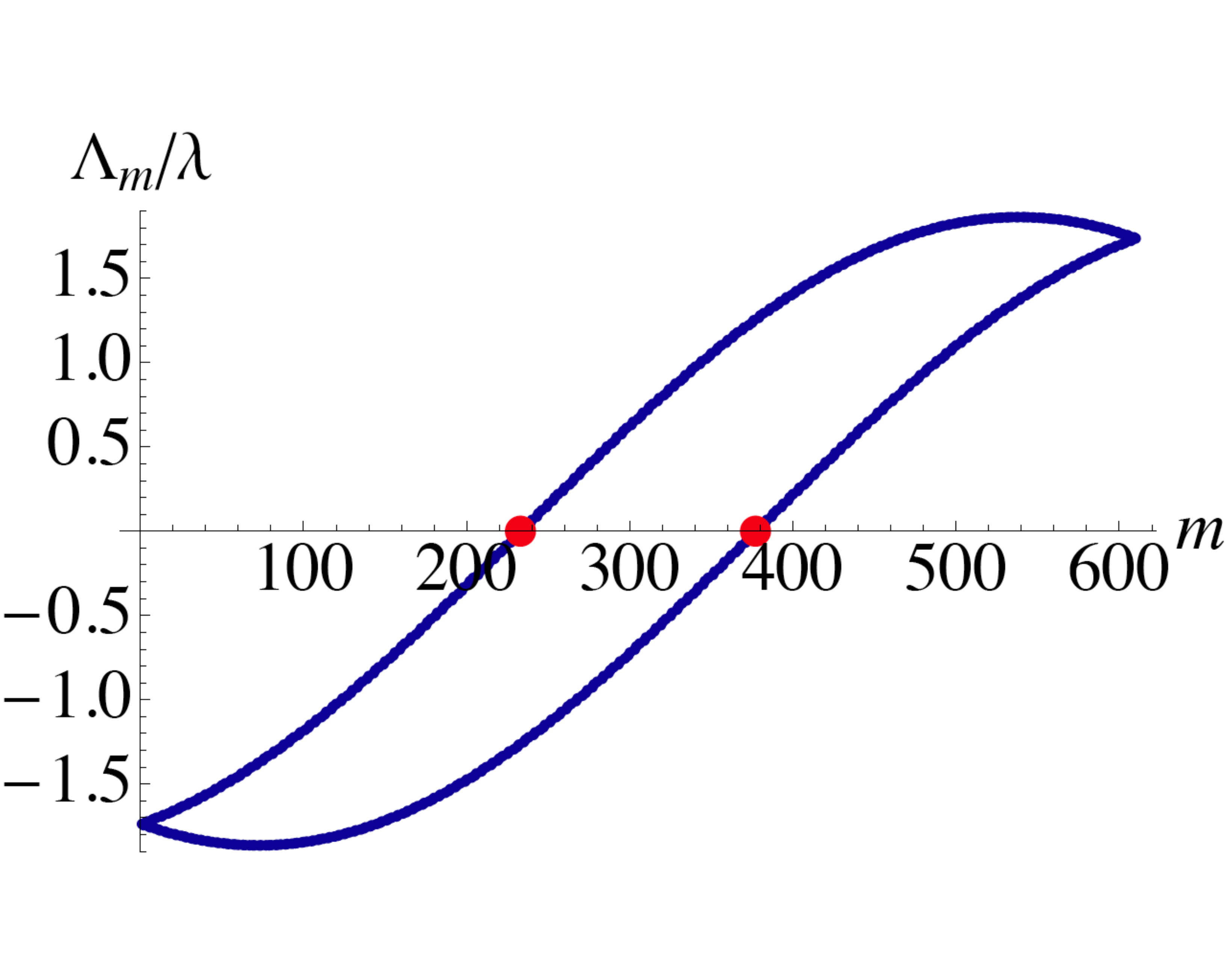}}\\
\subfigure[]{\includegraphics[width=1\linewidth]{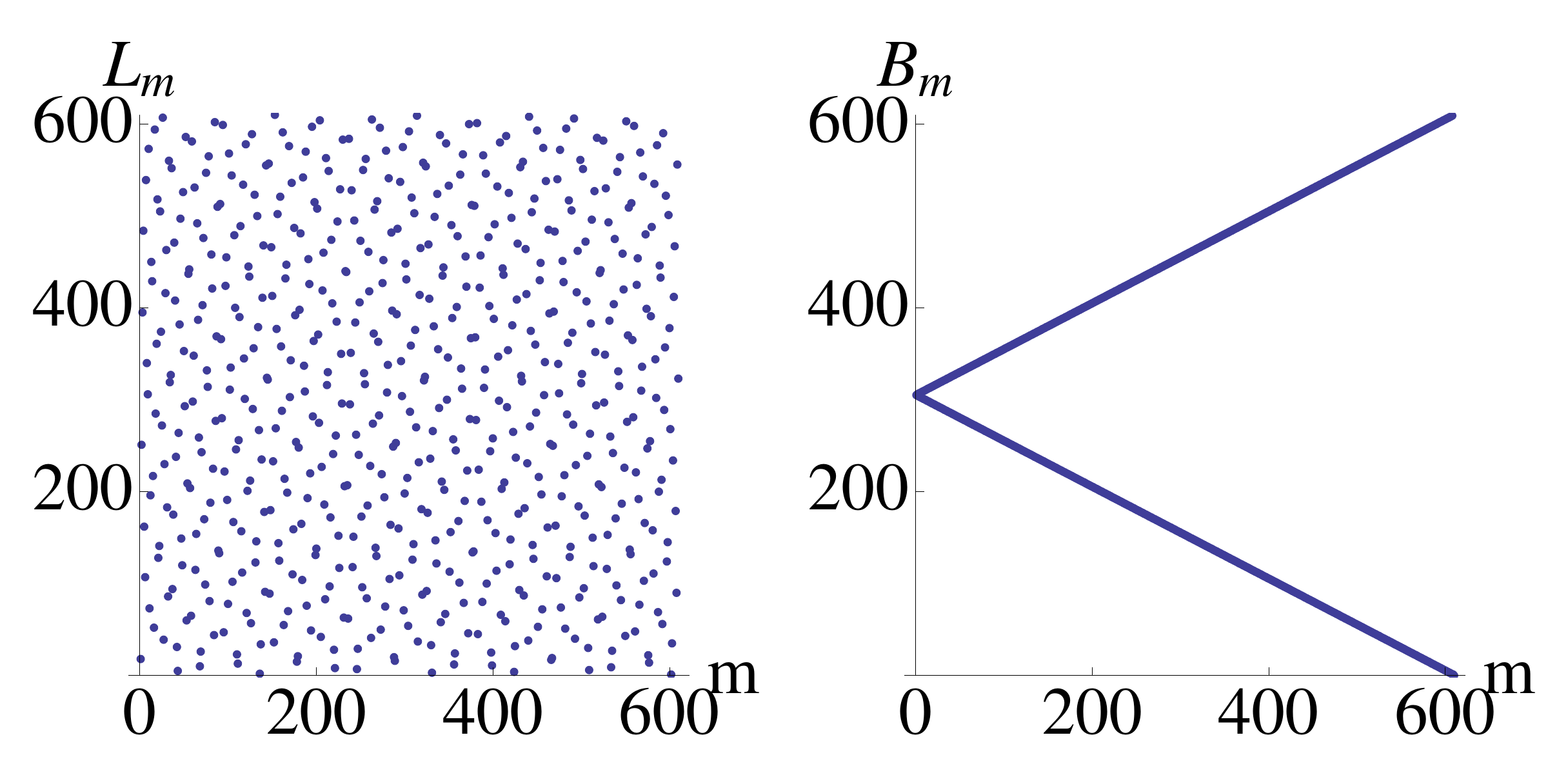}}
\caption{(Color online): Panel (a) shows numerically that at the two red points, $m_1=F_{M-1}$ and $m_2=F_{M-2}$, $\Lambda_m$  vanishes, so that degenerate perturbation theory is required.  Panel (b) shows  the relations of $L_m$, $B_m$ and $m$ (see text) for  $\phi=\frac{3\pi}{4}$. The relations between $L_{m}$ and $m$ are very sensitive to the value of $\phi$ while the relations between $B_{m}$ and $m$ are not sensitive at all.} \label{diffphi}
\end{figure}

When $|\Lambda_{m}|\ll1$, $|\beta|^{2}$ is comparable to $|\alpha|^{2}$ and   the states
are localized at the same two neighboring sites. Hence, they  will be referred as a {\it pair of dimerized states} ( The numerical results are shown in  Fig.\ref{diffphi} (a) ). The dimerized states are found by noticing that:
\begin{eqnarray}
 &&\cos(2\pi \sigma L_m+\phi)-\cos(2\pi\sigma(L_m+1)+\phi) \nonumber \\
 =&&2\sin\left(2\pi\sigma(L_m+1/2)+\phi\right)\sin(\pi\sigma)
\end{eqnarray}

So for the points that satisfy
$|\sin\left(2\pi\sigma(L_m+\frac{1}{2})+\phi\right)|\ll \frac{1}{\lambda}$,  dimerized sites exist at $L_{m}$ and $L_{m}+1$.
 Since
$\sigma=\frac{F_{M-1}}{F_M}$, it is clear that we can construct the
relation  $\text{Mod}[2\pi\sigma L_m+\phi,2\pi]\approx 2\pi\  \text{Mod}\left[ B_m/F_M,1\right]$, where
$m\in(1,2,\dots,F_M)$ are indexes of increasing energy level,
$L_m\in(1,2,\dots,F_M)$ are indexes of sites' positions and 
$B_m=\text{Mod}[L_m\times F_{M-1}+l,F_M]\in(0,1,2,\dots,F_M-1)$ are indexes
that we introduce for the convenience of discussion ($l$ is the closest integer to $\frac{\phi}{2\pi}F_{M}$). Using this relation  it can be shown  that $m\approx2|B_m-F_M/2|$ (we use $\approx $ here, since the exact expression for even and odd $F_M$ are slightly different). The relations of $m$,
$L_m$ and $B_m$ are plotted for the case of $F_M=610$ in Fig.\ref{diffphi} (b).

As we discussed, the dimerized states $L_m$ satisfy:
\begin{equation}
\sin(2\pi \sigma (L_m+1/2)+\phi)\approx 0
\end{equation} Writing  the above equation in terms of $B_m$, we get:
$$\sin\left(\frac{2\pi B_m}{F_M}+\pi\sigma\right)\approx0.$$
The solutions are $\frac{2B_m+F_{M-1}}{F_M}\approx 1\ \text{or}\ 2$. Therefore,
\begin{eqnarray}
B_{m_1}\approx (F_M-F_{M-1})/2\\
B_{m_2}\approx (2 F_M-F_{M-1})/2
\end{eqnarray}which correspond to:
\begin{eqnarray}
m_1\approx |2(B_{m_1}-F_{M/2})|\approx F_{M-1}\\
m_2\approx |2(B_{m_2}-F_M/2)|\approx F_{M-2}
\end{eqnarray}

A detailed analysis shows that the band  opens exactly at $m_1=F_{M-1}$ and $m_2=F_{M-2}$ regardless of   $F_{M}$ being   even or odd. This demonstrates, at perturbative level, the special  behavior of the return map at  irrational filling number  $F_{M-2}$ or
$F_{M-1}$(See Fig.\ref{returnirrational}).

In our calculations, we find out  those $L_m$ with the  lowest values of
$|\sin\left(2\pi\sigma(L_m+1/2)+\phi\right)|$ (avoiding double counting of $L_m$ in different pairs) and then apply to those points
the perturbation theory we discussed above.

Based on the latter considerations it is possible to  demonstrate that around  $m_1= F_{M-1}$ or $m_2=F_{M-2}$,  there exists a sequence of paired states,
 exhibiting the following relationships as $\lambda \rightarrow \infty$,

\begin{eqnarray*}
 \psi_{d}^{i,\tau}&=& \sqrt{A_{i}}  \delta_{j,L_{m_{\tau}-i}}+\sqrt{1-A_{i}} \delta_{j,L_{m_{\tau}-i}+1}\\
\psi_{u}^{i,\tau}&=& \sqrt{1-A_{i}} \delta_{j,L_{m_{\tau}-i}}-\sqrt{A_{i}} \delta_{j,L_{m_{\tau}-i}+1}
\label{pairedstates} 
\end{eqnarray*}with $i\ll F_{M}$ and $\tau=1,2$. These states determine the properties of the system when the Fermi energy is close to the major gaps,
i.e.  $\nu=\sigma$ or $\nu=\sigma^2$, and thus the main properties of the   band insulator phases.

\bibliography{./ref}
\end{document}